\documentclass[english,prd,a4paper,onecolumn,nofootinbib]{revtex4}
\pdfoutput=1
\usepackage{graphicx, bm, color, babel}
\usepackage{hyperref}
\usepackage{amsmath}
\usepackage{amssymb}
\usepackage{amsfonts}
\usepackage{dsfont}
\usepackage{color}

\def\be{\begin{equation}}
\def\ee{\end{equation}}
\def\bea{\begin{eqnarray}}
\def\eea{\end{eqnarray}}

\newcommand{\HH}{\mathcal{H}}

\newcommand\lsim{\mathrel{\rlap{\lower4pt\hbox{\hskip1pt$\sim$}}
        \raise1pt\hbox{$<$}}}

\newcommand{\Om}{\Omega}

\newcommand{\bn}{\mathbf{n}}
\newcommand{\bk}{\mathbf{k}}

\newcommand{\bx}{\mathbf{ x}}

\newcommand\spart{\;\raise1.0pt\hbox{$/$}\hskip-6pt\partial}
\newcommand\spartb{\;\overline{\raise1.0pt\hbox{$/$}\hskip-6pt
\partial}}

\newcommand{\B}{{\rm B}}
\newcommand{\F}{{\rm F}}
\newcommand{\w}[2]{w_{\bx_{#1}\bx_{#2}L_{#1} L_{#2}}}
\newcommand{\dn}{d\bar{n}}

\newcommand{\bv}{\mathbf{V}}

\newcommand{\norm}{a_N}

\begin{document}

\title{Measurement of the dipole in the cross-correlation function of galaxies}
\author{Enrique Gaztanaga$^1$}
\author{Camille Bonvin$^{2, 3}$}
\author{Lam Hui$^4$}
\affiliation{$^1$Institute of Space Sciences (IEEC-CSIC), Campus UAB, c/ can Magrans S/N 08193 Barcelona.\\
$^2$CERN, Theory Division, 1211 Geneva, Switzerland.\\
$^3$D\'epartement de Physique Th\'eorique and Center for Astroparticle Physics (CAP), University of Geneva,
24 quai Ernest Ansermet, CH-1211 Geneva, Switzerland.\\
$^4$Institute for Strings, Cosmology and Astroparticle Physics
and Department of Physics, 
Columbia University, New York, NY 10027, U.S.A.\\
{\rm E-mail: gazta@ice.cat, camille.bonvin@unige.ch, lhui@astro.columbia.edu}}  \vspace*{0.2cm}

\date{\today}

\begin{abstract} 

It is usually assumed that in the linear regime the two-point correlation function of galaxies contains only a monopole, quadrupole and hexadecapole. Looking at cross-correlations between different populations of galaxies, this turns out not to be the case. In particular, the cross-correlations between a bright and a faint population of galaxies contain also a dipole. In this paper we present the first attempt to measure this dipole. We discuss the four types of effects that contribute to the dipole: relativistic distortions, evolution effect, wide-angle effect and large-angle effect. We show that the first three contributions are intrinsic anti-symmetric contributions that do not depend on the choice of angle used to measure the dipole. On the other hand the large-angle effect appears only if the angle chosen to extract the dipole breaks the symmetry of the problem. We show that the relativistic distortions, the evolution effect and the wide-angle effect are too small to be detected in the LOWz and CMASS sample of the BOSS survey. On the other hand with a specific combination of angles we are able to measure the large-angle effect with high significance. We emphasise that this large-angle dipole does not contain new physical information, since it is just a geometrical combination of the monopole and the quadrupole. However this measurement, which is in excellent agreement with theoretical predictions, validates our method for extracting the dipole from the two-point correlation function and it opens the way to the detection of relativistic effects in future surveys like e.g. DESI.
\end{abstract}

\maketitle

\section{Introduction}

The two-point correlation function of galaxies is a powerful cosmological probe. It has been measured with increasing precision in various surveys over the last 50 years. In those measurements, the correlation function is implicitly assumed to be symmetric under the exchange of the galaxies in the pair $\langle \Delta(\bx_i)\Delta(\bx_j)\rangle=\langle \Delta(\bx_j)\Delta(\bx_i)\rangle$, where $\Delta(\bx_i)$ represents the fractional over-density of galaxies in pixel $i$. In redshift-space, the correlation function can be expressed as a function of the pixels separation $d_{ij}$, the redshift of the pair, and the angle that the pair makes with the observer's line-of-sight $\gamma_{ij}$. The symmetry of the correlation function implies that it depends only on even powers of $\cos\gamma_{ij}$. More precisely, in the distant-observer approximation, the two-point correlation function can simply be written as the sum of a monopole, a quadrupole and an hexadecapole~\cite{Kaiser:1987qv, 1989MNRAS.236..851L, 1992ApJ...385L...5H}. The measurement of these multipoles constitutes one of the great success of large-scale structure surveys since it provides a direct measurement of the growth rate of cosmological perturbations $f$ (see e.g. \cite{Hawkins:2002sg, Zehavi:2004zn, Guzzo:2008ac, Cabre:2008sz, Song:2010kq, Samushia:2013yga}).

Recently it has been shown that if one splits the galaxies into different populations (with e.g. different luminosities or colours), the form of the two-point correlation function becomes more complex. In particular, the cross-correlations between a bright ($\B$) and a faint ($\F$) population of galaxies acquire an anti-symmetric part~\cite{McDonald:2009ud, Yoo:2012se, Croft:2013taa, Bonvin:2013ogt}:  $\langle \Delta_\B(\bx_i)\Delta_\F(\bx_j)\rangle\neq\langle \Delta_\F(\bx_i)\Delta_\B(\bx_j)\rangle$. The goal of this paper is to measure, for the first time, this anti-symmetric part in the correlation function.

As shown in~\cite{Bonvin:2013ogt, Bonvin:2014owa}, there are three types of effects that generate an anti-symmetry in the correlation function~\footnote{Note that here we do not consider possible primordial anti-symmetries in the correlation function generated for example by a primordial vector field~\cite{Dai:2015wla}.}. The first contribution is due to {\it relativistic distortions} in the number count of galaxies $\Delta$. This observable is indeed not only affected by density fluctuations and by the gradient of the velocity, as assumed in the standard redshift-space distortion expression, but also by various contributions of the gravitational potentials and of the peculiar velocity~\cite{Yoo:2009au, Yoo:2010ni, Bonvin:2011bg, Challinor:2011bk, Jeong:2011as}. Among all these effects, one can show that the following combination of gravitational redshift and Doppler terms generates an anti-symmetry in the correlation function
\begin{align}
\label{Deltarel}
\Delta^{\rm anti}(\bx, z)=&\frac{1}{\HH}\partial_r\Psi + \frac{1}{\HH}\dot{\bv}\cdot\bn+\left[1-\frac{\dot\HH}{\HH^2}-\frac{2}{r\HH}+5s\left(1-\frac{1}{r\HH} \right)\right]\bv\cdot\bn\, ,
\end{align}
where $\Psi$ is the time-component of the metric~\footnote{We use here the following convention for the metric $ds^2=a^2\big[-(1+2\Psi)d\eta^2+(1-2\Phi)\delta_{ij}dx^i dx^j \big]$, where $a$ is the scale factor and $\eta$ denotes conformal time.}, $\bv$ is the peculiar velocity, $\bn$ is the observed direction, $\HH$ is the conformal Hubble parameter, $r$ is the conformal distance to the source and $s$ is the slope of the luminosity function. A dot denotes derivative with respect to conformal time $\eta$. The terms in Eq.~\eqref{Deltarel} contain all one gradient of the potential, which is responsible for the anti-symmetry~\footnote{Gravitational lensing also generates an anti-symmetry in the two-point function. However, as shown in~\cite{Bonvin:2013ogt} this contribution is always significantly smaller than the terms in Eq.~\eqref{Deltarel} and it can safely be neglected.}. Note that in Fourier space, these terms have the particularity to generate an imaginary part to the power spectrum~\cite{McDonald:2009ud, Yoo:2012se}.

The second source of anti-symmetry in the correlation function comes from the {\it evolution} of the bias and of the growth rate. This evolution --which differs for different populations of galaxies-- generates a systematic asymmetry between e.g. the number of faint galaxies in front and behind a bright galaxy. 

Finally, the third effect generating an anti-symmetry in the two-point function is the {\it wide-angle} effect. Beyond the distant-observer approximation, the line-of-sight to the two galaxies in the pair are not parallel. This generates corrections to the standard Kaiser expression for the monopole, quadrupole and hexadecapole (see e.g. \cite{Hamilton:1997zq, Szalay:1997cc, Szapudi:2004gh, Papai:2008bd, Raccanelli:2010hk, Samushia:2011cs, Montanari:2012me, Bertacca:2012tp, Yoo:2013zga, 2016JCAP...01..048R}), but it also induces an additional anti-symmetric contribution to the two-point correlation function~\cite{Bonvin:2013ogt}. Unsurprisingly, the form of the wide-angle effect depends on the choice of angle used to extract the multipoles. We explore two different choices commonly used in large-scale structure surveys (see Figure~\ref{fig:angles}) and show that with one of them the wide-angle dipole is directly related to the fact that redshift-space distortions depend on {\it two} line-of-sights. With the other choice of angle on the other hand, another contribution to the dipole appears, which follows from the fact that the coordinate system itself breaks the symmetry of the correlation function. We call this extra contribution {\it large-angle} effect~\footnote{Note that some authors, e.g. \cite{2016JCAP...01..048R}, do not make the distinction between wide-angle and large-angle effects and simply call wide-angle effect all contributions to the multipoles that are suppressed by one or more powers of $d/r$ where $d$ is the comoving separation between galaxies and $r$ is the comoving distance to the pair.}.

The anti-symmetry generated by the relativistic, evolution, wide-angle and large-angle effects can be expanded in odd multipoles of $\cos\gamma_{ij}$~\cite{Bonvin:2013ogt, Raccanelli:2013dza}. The dominant contribution is a dipole, simply proportional to $\cos\gamma_{ij}$. In this paper, we show how to measure this dipole in the two-point correlation function of galaxies. Note that this dipole is completely different from the kinematic dipole due to our motion with respect to the frame of the Cosmic Microwave Background~\cite{Kogut:1993ag}. The kinematic dipole is a dipole around the observer, whereas the dipole we are measuring in this paper is a dipole around each of the bright galaxies in the survey.  

The remainder of the paper is organised as follow: in Section~\ref{sec:kernel} we present the method used to extract the dipole from the two-point correlation function of galaxies. We show that depending on the choice of kernel, we are sensitive to different combinations of terms. In Section~\ref{sec:measure} we apply our method to the BOSS LOWz and CMASS samples. We show that the contributions due to the relativistic distortions, evolution and wide-angle effect are too small to be detected in these samples. We calculate the signal-to-noise of these contributions and show that it is of the order of 0.2. On the other hand we show that the large-angle effect is large enough to be measured in the LOWz and CMASS samples. As explained in detail in Section~\ref{sec:wide-angle} this large-angle effect does not contain new statistical information since it is simply a geometrical effect related to a specific choice of angle that breaks the symmetry of the situation. However this detection is interesting for two reasons: first it shows that if one wants to measure relativistic effects in future surveys it is crucial \textit{not} to use this choice of angle, which artificially introduces anti-symmetries in the correlation function. Second, this detection validates our method to measure a dipole in galaxy surveys, since it is in very good agreement with our theoretical prediction for the large-angle effect. We then forecast the detectability of relativistic effects with a future survey like DESI and show that in this case the cumulative signal-to-noise reaches 7.4.
In Section~\ref{sec:zshell}, we compare our method with the estimator $z_{\rm shell}$ presented in~\cite{Croft:2013taa} and show that a simple expression relates the two estimators. We also compare our method with the measurement of gravitational redshift in clusters~\cite{Wojtak:2011ia, Sadeh:2014rya} and we discuss why relativistic effects are more challenging to measure in large-scale structure. Note that in this paper we are interested in galaxy correlations at large scales. We work therefore in linear perturbation theory, where the non-linear effects described in~\cite{2013PhRvD..88d3013Z, Kaiser:2013ipa} are negligible. Finally in Section~\ref{sec:1pop} we show that the large-angle dipole can also be measured with a single population of galaxies if we choose the appropriate kernel. We conclude in Section~\ref{sec:conclusion}.

\section{Extracting the dipole}
\label{sec:kernel}

As mentioned above, to measure an anti-symmetry in the two-point correlation function we need in general more than one population of galaxies. In the following, we split therefore the galaxies into a bright population and a faint population based on their luminosity. In a companion paper~\cite{Bonvin:2015kuc}, we consider the more general case of multiple populations of galaxies with different luminosities and we show how the dipole can be optimally extracted in this case. Here we restrict ourselves to two populations and we adjust the threshold between the bright and the faint population to obtain a similar number of galaxies in each sample. 

For each pixel $i$ in the sky, we count the number of bright galaxies $n_\B(\bx_i)$ and the number of faint galaxies $n_\F(\bx_i)$. 
We construct the over-density of galaxies
\be
\delta n_\B(\bx_i)=n_\B(\bx_i)-\dn_\B \quad \mbox{and}\quad \delta n_\F(\bx_i)=n_\F(\bx_i)-\dn_\F\,,
\ee
where $\dn_\B$ and $\dn_\F$ denote the mean number of galaxies per pixel. Note that these quantities depend on the size of the pixels. 
We combine the two populations of galaxies with a kernel $w_{\bx_i \bx_j L_i L_j}$ designed to extract the dipole
\be
\label{estimator}
\hat\xi=\sum_{ij}\sum_{L_i L_j=\B, \F} w_{\bx_i \bx_j L_i L_j}\delta n_{L_i}(\bx_i)\delta n_{ L_j}(\bx_j)\, .
\ee
To isolate the dipole from the other multipoles, the kernel $w_{\bx_i \bx_j L_i L_j}$ must be proportional to the cosine of the angle $\gamma_{ij}$ that the pair of pixels makes with the line-of-sight. It must also depend on the respective luminosities of the pixels, $L_i$ and $L_j$. A simple way of constructing a kernel with these properties is
\be
\label{kernel2}
\w{i}{j}=\norm \cos\gamma_{ijL_i L_j}\delta_K(d_{ij}-d)\, ,
\ee
where 
\be
\label{gamma}
\gamma_{ijL_i L_j}=\left\{\begin{array}{c}\gamma_{ij} \hspace{0.8cm}\mbox{if}\hspace{0.8cm} L_i>L_j\,,\\
\gamma_{ji} \hspace{0.8cm}\mbox{if}\hspace{0.8cm} L_i<L_j\, . \end{array}\right.
\ee
Here $\norm$ is a normalisation factor (which depends on the pair separation $d=d_{ij}$)
\be
\label{norm}
\norm=\frac{3}{8\pi}\big(N_{\rm tot}\ell_p d^2 \bar N \bar n_\B\bar n_\F\big)^{-1}\, ,
\ee
with $N_{\rm tot}$ the total number of galaxies, $\ell_p$ the length of a cubic pixel, $\bar N$ the mean number density of galaxies, and $\bar n_\B$ and $\bar n_\F$ the fractional number of bright and faint galaxies. Note that this kernel is slightly different from the one chosen in~\cite{Bonvin:2015kuc}, which we discuss in Appendix~\ref{app:kernel}.

\begin{figure}
\centering
\includegraphics[width=0.45\textwidth]{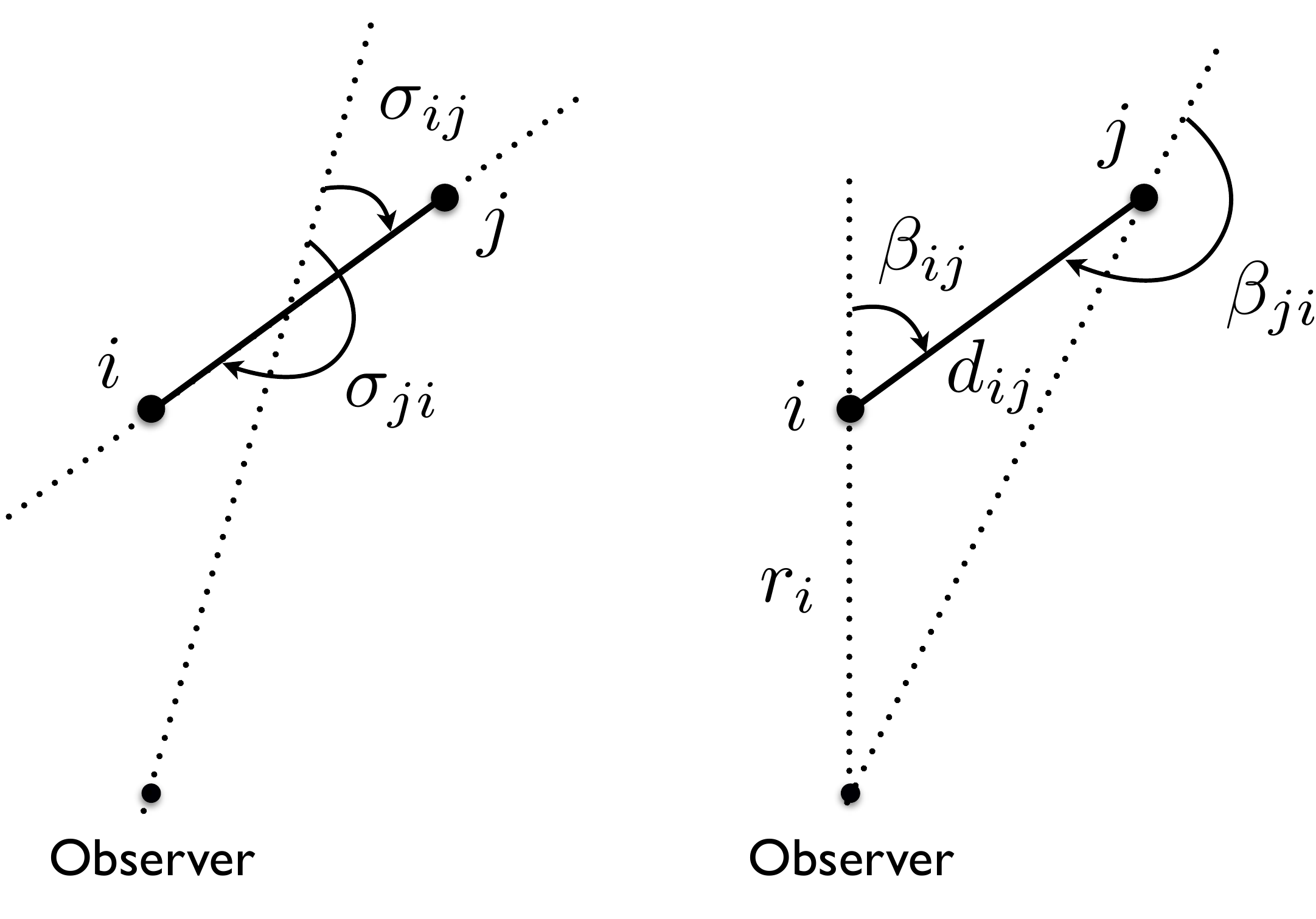}
\caption{\label{fig:angles} Representation of the two cases we consider for the line-of-sight angle: $\sigma_{ij}$ (left panel) and $\beta_{ij}$ (right panel). In Section~\ref{sec:measure}, we measure the dipole with  $\mu_{12}\equiv\cos\sigma_{ij}$ and with $\mu_1=\cos\beta_{ij}$. We also construct an estimator based on the difference between the two angles: $\cos\beta_{ij}-\cos\sigma_{ij}$.}
\end{figure}

Beyond the distant-observer approximation, there are different ways of choosing the angle between the pair of galaxies and the observer's line-of-sight. In Figure~\ref{fig:angles} we show two possible choices: $\gamma_{ij}=\sigma_{ij}$, i.e. the angle between the median and the vector connecting the median to $j$; and $\gamma_{ij}=\beta_{ij}$, i.e. the angle between the direction of the pixel $i$ and the vector connecting $i$ to $j$. The first choice $\sigma_{ij}$ has the advantage that it places the two pixels $i$ and $j$ on the same footing: it is therefore by construction the angle that respects the most the symmetry of the problem. The angle $\beta_{ij}$ on the other hand seems a natural choice to measure the relativistic corrections: it places the bright galaxy at the centre of the coordinate system and tells us how the amplitude of the correlation varies around the bright galaxy. An effect like gravitational redshift is expected to create a dipole modulation around each bright galaxy due to the difference in gravitational potential between the bright galaxy sitting at the bottom of the gravitational potential and the faint galaxies around. To measure such an effect it seems therefore natural to use the angle defined with respect to the bottom of the potential. This is for example the angle used in~\cite{Wojtak:2011ia, Sadeh:2014rya} to measure gravitational redshift in clusters and in~\cite{Croft:2013taa} to construct the estimator $z_{\rm shell}$ aimed at measuring gravitational redshift in large-scale structure.

\subsection{Relativistic and evolution dipole}

Even though the angle $\beta_{ij}$ seems more intuitive to measure the relativistic dipole, we can show that at lowest order in $d/r$ the form of the dipole is exactly the same with $\beta_{ij}$ or $\sigma_{ij}$. It is only at the order $(d/r)^3$ that the two angles give a different result. For the scales used in this work these differences are negligible. Similarly, the evolution dipole is independent of the choice of angle at leading order. In the continuous limit the mean of these contributions can be written as (see \cite{Bonvin:2013ogt} for more detail)
\begin{align}
\langle\hat\xi^{\rm rel}\rangle=&(b_\B-b_\F)\left(\frac{\dot\HH}{\HH^2}+\frac{2}{r\HH} \right)\frac{\HH}{\HH_0}\frac{f}{2\pi^2}\int dk k \HH_0 P(k, \bar z)j_1(kd)\,  ,\label{rel}\\
\langle\hat\xi^{\rm evol}\rangle=&\frac{d}{r}\Bigg\{\frac{r}{6}\Big[(b_\B-b_\F)f'-f(b'_\B-b'_\F)\Big]\left(C_0(d)-\frac{4}{5}C_2(d) \right)+\frac{r}{2}\big(b_\B b'_\F-b'_\B b_\F\big)C_0(d)\Bigg\} \, ,\label{evol}
\end{align}
where $f$ is the growth rate, $b_\B$ and $b_\F$ represent the bias of the bright and faint population respectively and a prime denotes a derivative with respect to the conformal distance $r$. Here $P(k, \bar z)$ is the density power spectrum at the mean redshift of the survey:
\be
\langle \delta(\bk, \bar z)\delta(\bk', \bar z) \rangle=(2\pi)^3 P(k, \bar z)\delta_D(\bk+\bk')\, ,
\ee
and 
\be
C_\ell(d)=\frac{1}{2\pi^2}\int dk k^2 P(k, \bar z) j_\ell(kd)\, ,\quad \ell= 0, 2\, .
\ee
Note that contrary to the angular power spectrum, which is very sensitive to the size of the redshift bin within which the pairs of pixels $i, j$ are averaged~\cite{Bonvin:2011bg, DiDio:2013sea}, the two-point correlation function $\hat\xi$ is almost insensitive to the binning procedure~\cite{Bonvin:2014owa}. Therefore in Eqs.~\eqref{rel} and~\eqref{evol} we approximate the average over redshift by the value of the correlation function at the mean redshift of the survey $\bar z$. 

\subsection{Wide-angle and large-angle dipole}
\label{sec:wide-angle}

Contrary to the relativistic and evolution dipole, the wide-angle and large-angle dipole depend strongly on the choice of angle (note that this dependence is discussed in detail in~\cite{2016JCAP...01..048R} for the case of one population of galaxies). Using the median angle $\sigma_{ij}$, we find for the mean of the wide-angle effect
\begin{align}
\langle \hat\xi^{\rm wide\, \sigma} \rangle=&\norm\dn_{\B}\dn_{\F}\sum_{ij}\cos\sigma_{ij}\delta_K(d_{ij}-d) \Big[\langle \Delta_\B(\bx_i)\Delta_\F(\bx_j)\rangle-\langle \Delta_\B(\bx_j)\Delta_\F(\bx_i)\rangle\Big]^{\rm wide}\, ,\label{xisigma}
\end{align}
where we have used that $\cos\sigma_{ji}=-\cos\sigma_{ij}$ (see Figure~\ref{fig:angles}). In~\cite{Bonvin:2013ogt} we calculated the term in bracket as a function of the angle $\beta_{ij}$. Here we need to rewrite it in terms of $\sigma_{ij}$. The calculation is presented in Appendix~\ref{app:wideangle}. We find in the continuous limit
\be
\label{wide1}
\langle \hat\xi^{\rm wide\, \sigma} \rangle=\frac{2f}{5}(b_\B-b_\F)\frac{d}{r}C_2(d)\, .
\ee

\begin{figure}
\centering
\includegraphics[width=0.48\textwidth]{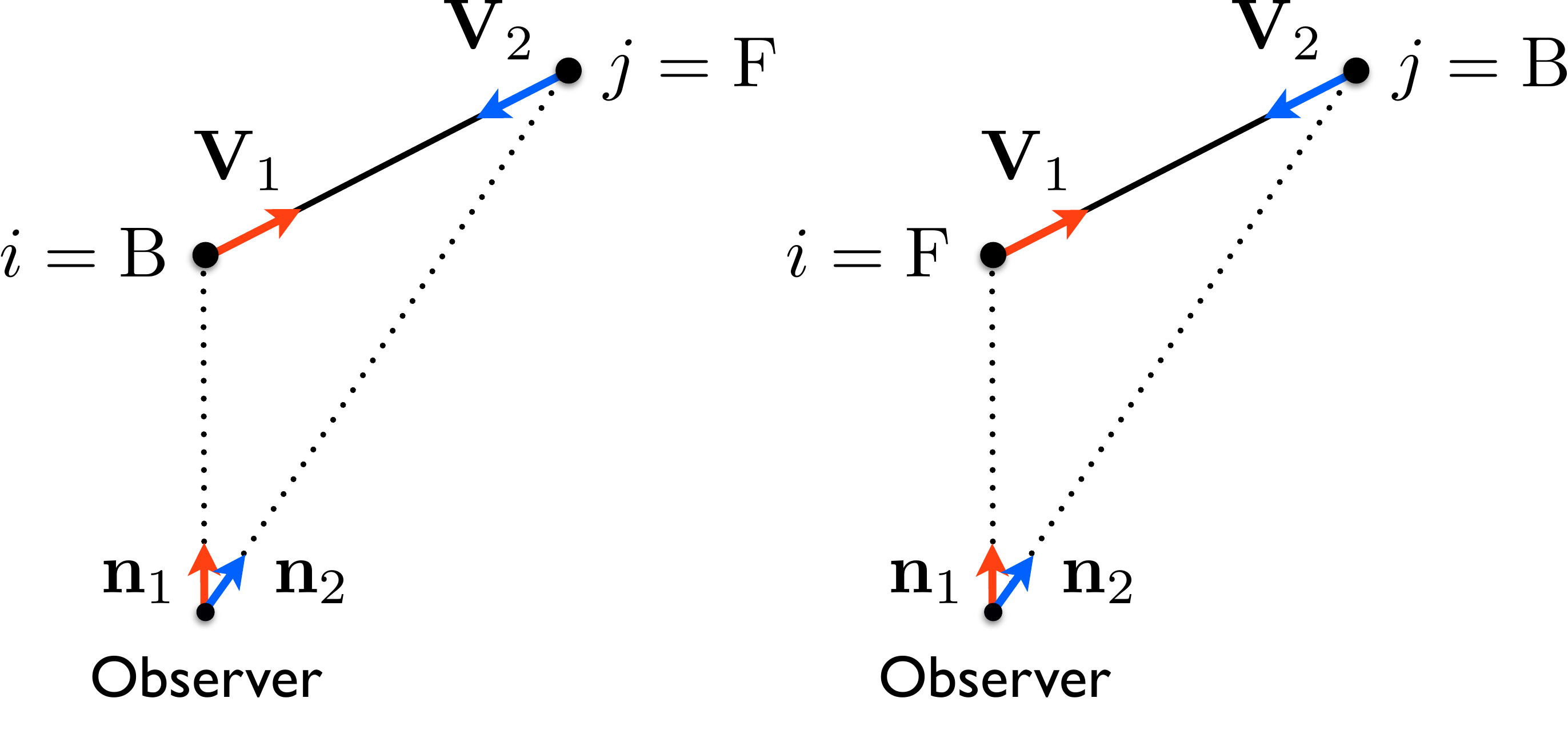}
\caption{\label{fig:wideangle} Contribution to the wide-angle dipole~\eqref{wide1} from the correlation between density and redshift-space distortion. In the flat-sky we have $\bn_1=\bn_2=\bn$ so that $|\bv_1\cdot \bn|=|\bv_2\cdot \bn|$ and the contributions from the left and right panels cancel. In the full sky however $\bn_1\neq\bn_2$ and the two panels combine to give~\eqref{wide1}.}
\end{figure}

The physical origin of this wide-angle effect can be understood by looking at Figure~\ref{fig:wideangle}. From the left panel, we see that the correlation between redshift-space distortion and density generates a contribution proportional to $|\bv_1\cdot \bn_1|$ weighted by the bias of the faint galaxy plus a contribution proportional to $|\bv_2\cdot \bn_2|$ weighted by the bias of the bright galaxy. To these contributions we must, according to Eq.~\eqref{xisigma}, subtract the contributions from the right panel of Figure~\ref{fig:wideangle}. In this case, the contribution proportional to $|\bv_1\cdot \bn_1|$ is weighted by the bias of the bright galaxy and the contribution proportional to $|\bv_2\cdot \bn_2|$ is weighted by the bias of the faint galaxy. In the distant-observer approximation we have $\bn_1=\bn_2=\bn$. Since on average the galaxies tend to attract each other, we have then $|\bv_1\cdot \bn|=|\bv_2\cdot \bn|$ so that the contributions from the left and right panels of Figure~\ref{fig:wideangle} cancel. At large scale however, when the distant-observer approximation breaks down $\bn_1\neq\bn_2$. The contributions from the right and left panels of Figure~\ref{fig:wideangle} are not exactly the same and they give rise to Eq.~\eqref{wide1}.

If instead of using $\sigma_{ij}$ to measure the dipole we use $\beta_{ij}$ we obtain
\begin{align}
\label{xiw2beta}
\langle \hat\xi^{\rm wide\, \beta} \rangle=\norm&\dn_{\B}\dn_{\F}\sum_{ij}\Big[\langle \Delta_\B(\bx_i)\Delta_\F(\bx_j)\rangle\cos\beta_{ij}
+\langle \Delta_\B(\bx_j)\Delta_\F(\bx_i)\rangle\cos\beta_{ji}\Big]^{\rm wide}\delta_K(d_{ij}-d)\, .
\end{align}
From Figure~\ref{fig:angles} we see that at large separation $\cos\beta_{ij}\neq -\cos\beta_{ji}$. In the full sky we have
\begin{align}
\cos\beta_{ji}&=-\frac{r_i\cos\beta_{ij}+d_{ij}}{\sqrt{r_i^2+2r_id_{ij}\cos\beta_{ij}+d^2_{ij}}}
= -\cos\beta_{ij}-\frac{d_{ij}}{r_i}\left(1-\cos^2\beta_{ij}\right)+\mathcal{O}\left( \frac{d_{ij}}{r_i}\right)^2\, . \label{betaangle}
\end{align}
Inserting this into Eq.~\eqref{xiw2beta} we obtain at lowest order in $d_{ij}/r_i$
\begin{align}
\langle \hat\xi^{\rm wide\, \beta} \rangle=\norm\dn_{\B}\dn_{\F}\sum_{ij}&\Bigg\{\Big[\langle \Delta_\B(\bx_i)\Delta_\F(\bx_j)\rangle-\langle \Delta_\B(\bx_j)\Delta_\F(\bx_i)\rangle\Big]^{\rm wide}
\cos\beta_{ij}\\
&-\langle \Delta_\B(\bx_j)\Delta_\F(\bx_i)\rangle\frac{d_{ij}}{r_i}\left(1-\cos^2\beta_{ij}\right)\Bigg\}\delta_K(d_{ij}-d)\, .\nonumber
\end{align}
The term in the first line gives rise to Eq.~\eqref{wide1} (since the bracket is already of order $d_{ij}/r$, we can replace $\beta_{ij}$ by $\sigma_{ij}$ in the first line). The term in the second line however is a novel contribution, specific to this particular choice of angle. To differentiate this contribution from the one in Eq.~\eqref{wide1} we call it the {\it large-angle} contribution~\footnote{Note that this large-angle contribution also affects the monopole and quadrupole and it is sometimes included into the wide-angle contribution (see e.g.~\cite{2016JCAP...01..048R}).}. 
In the continuous limit, we obtain
\be
\label{wide2}
\langle \hat\xi^{\rm wide\, \beta} \rangle=\langle \hat\xi^{\rm wide\, \sigma} \rangle + \langle \hat\xi^{\rm large} \rangle\, ,
\ee
where
\begin{align}
\label{large}
\langle \hat\xi^{\rm large} \rangle=& -\frac{d}{r}\Bigg\{ \left[b_\B b_\F+(b_\B+b_\F)\frac{f}{3}+\frac{f^2}{5} \right]C_0(d)+\frac{1}{5}\left[(b_\B+b_\F)\frac{2f}{3}+\frac{4f^2}{7} \right]C_2(d)\Bigg\}\, .
\end{align}
The large-angle dipole~\eqref{large} has a different physical origin than the wide-angle dipole~\eqref{wide1}. As explained above, the wide-angle dipole is due to the fact that the two-point function is intrinsically anti-symmetric in the full sky due to the difference between the line-of-sight to the bright and to the faint galaxies. The large-angle dipole on the other hand is due to the fact that the angle we are using to measure the dipole breaks explicitly the symmetry of the situation. From this we understand already that the large-angle dipole exists also in the case where we have only {\it one} population of galaxies. In Section~\ref{sec:1pop} we will show how this term can be measured in that case. 

\begin{figure}
\centering
\includegraphics[width=0.47\textwidth]{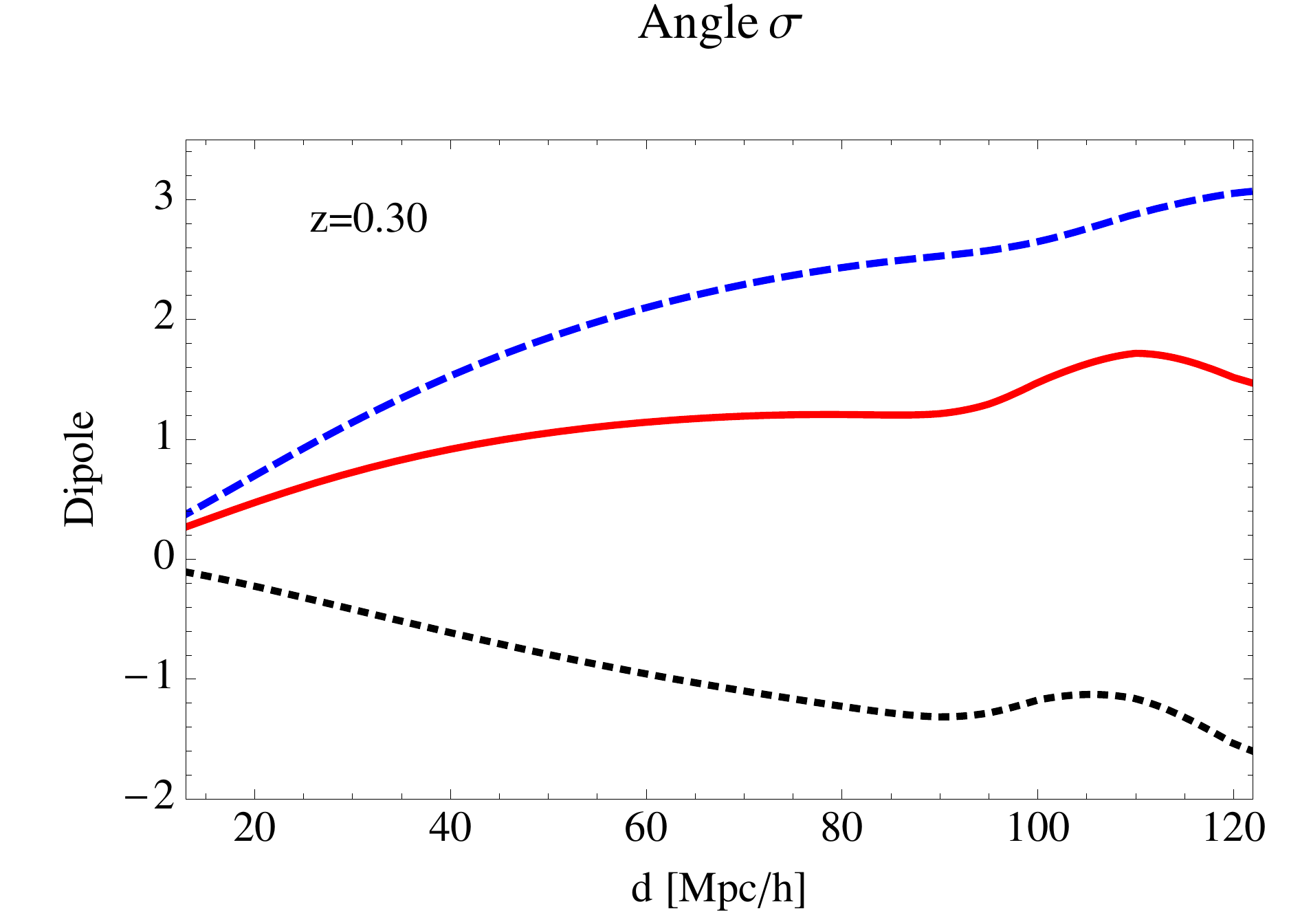}\hspace{0.8cm}\includegraphics[width=0.47\textwidth]{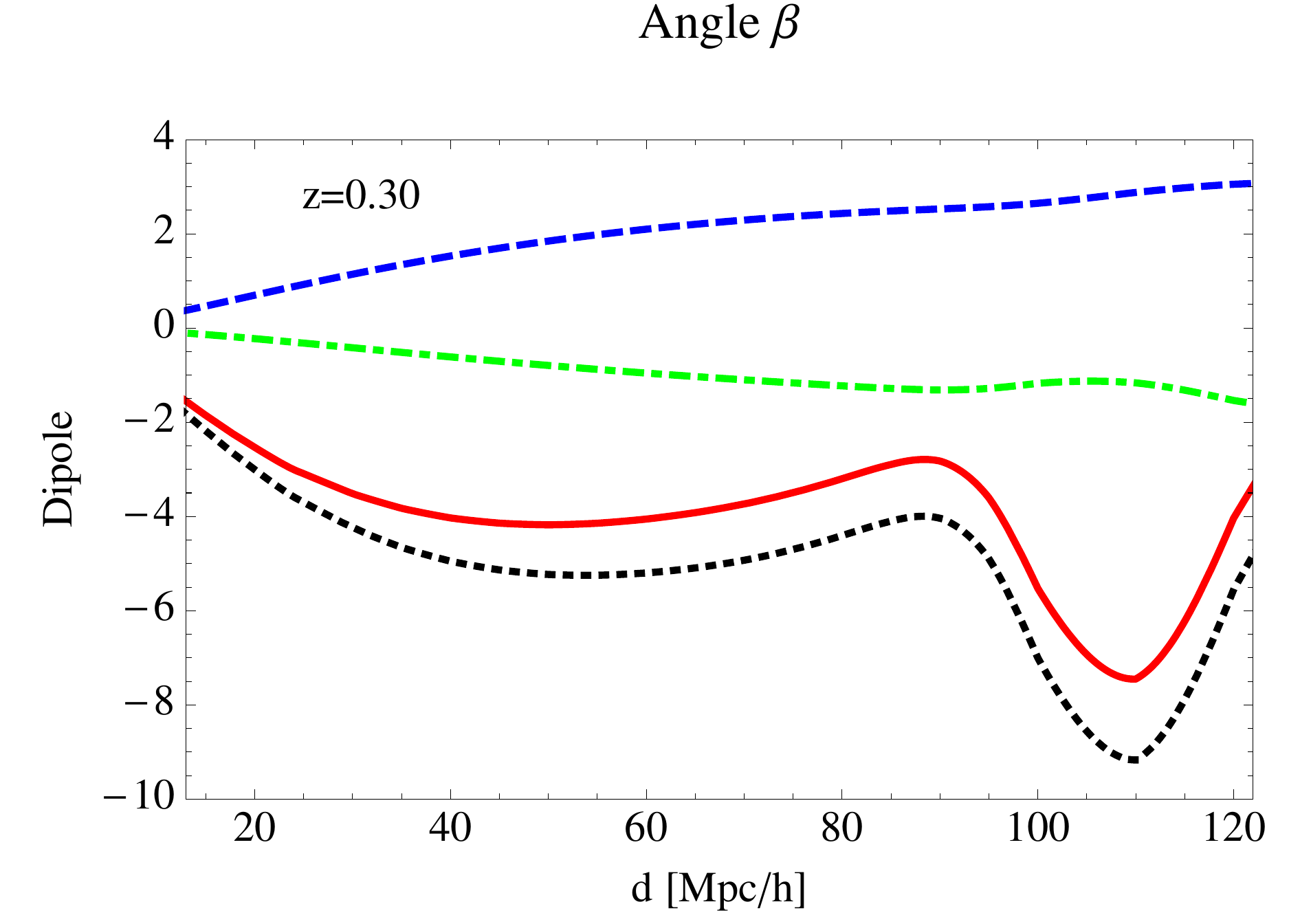}\\
\includegraphics[width=0.47\textwidth]{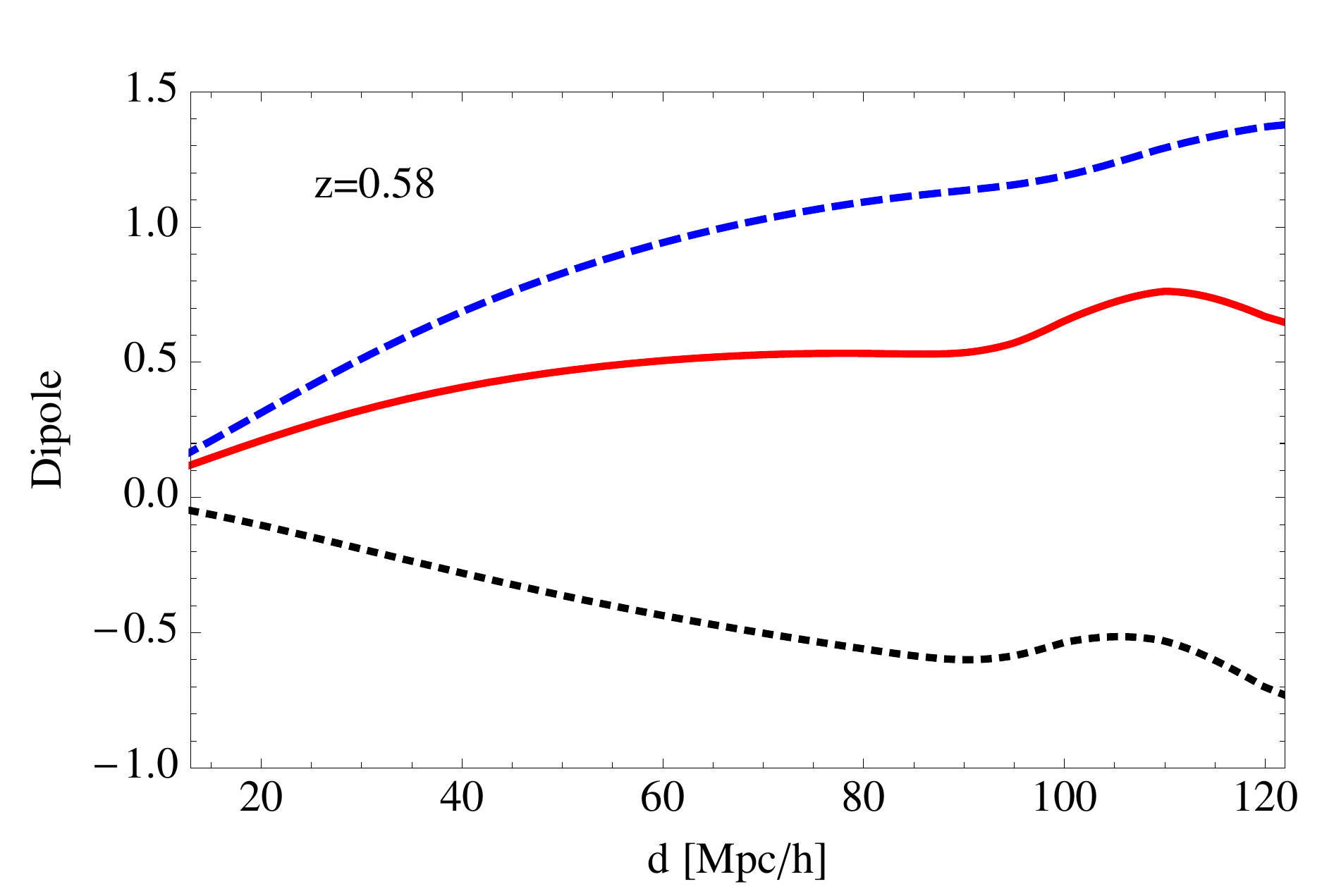}\hspace{0.8cm}\includegraphics[width=0.47\textwidth]{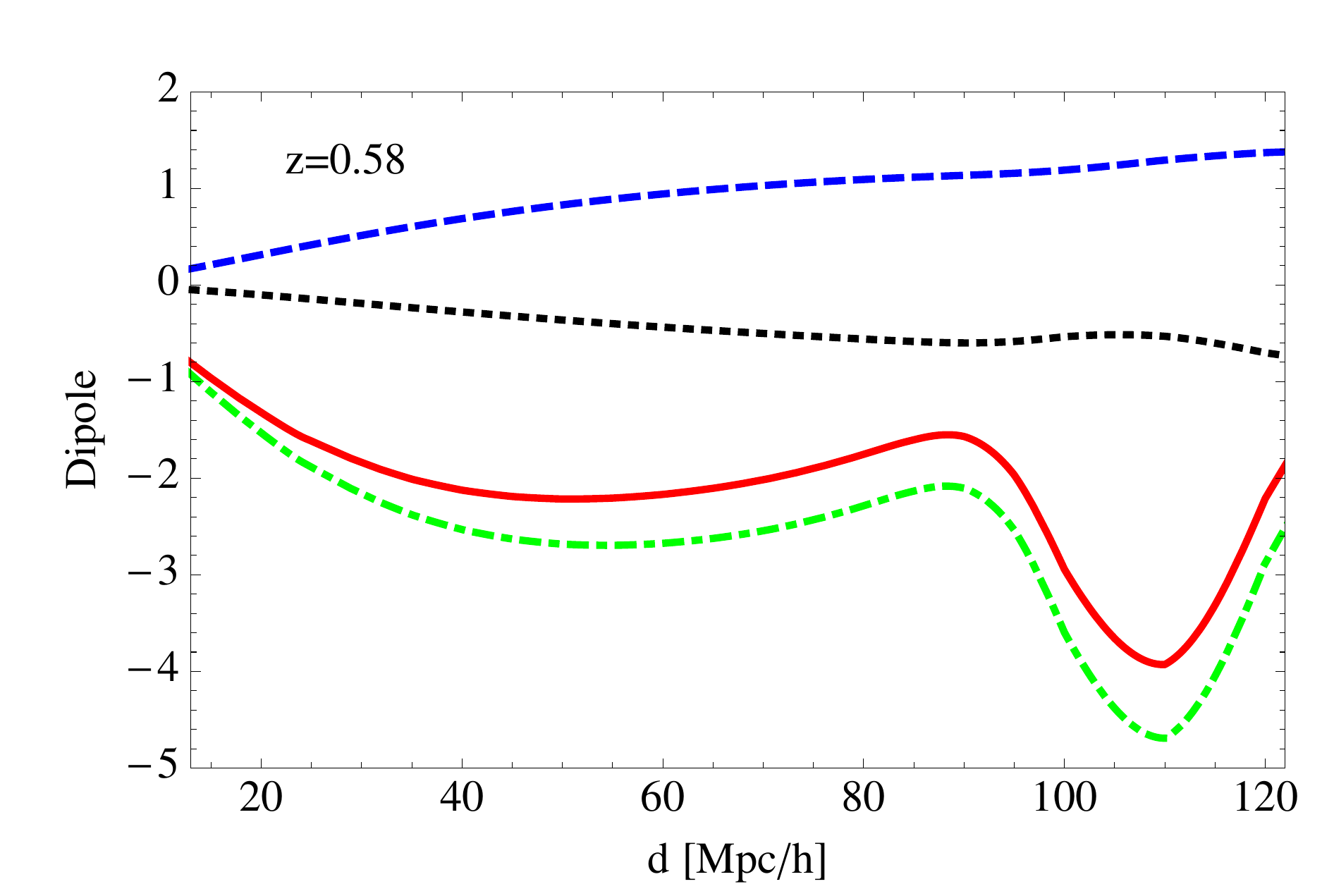}
\caption{\label{fig:predictions} Predictions for the observed dipole at $z=0.30$ and $z=0.58$, multiplied by $d^2$, and plotted as a function of separation. In the left panels we show the dipole measured with $\mu_1\equiv\cos\sigma$ and in the right panels with $\mu_{12}\equiv\cos\beta$. The blue dashed line is the relativistic dipole, the black dotted line is the wide-angle dipole, the green dot-dashed line is the large-angle dipole and the red solid line is the total. The dipole due to evolution is much smaller than the other contributions and we have therefore neglected it. Note that the scale on the y-axis is different for each plot.}
\end{figure}

In Figure~\ref{fig:predictions} we compare the different dipole contributions at $z=0.30$ and $z=0.58$ using the angle $\sigma$ and $\beta$. These contributions depend on the bias of the bright and faint populations. We use the values measured in Section~\ref{sec:measure} to evaluate them. The dipole due to evolution depends furthemore on the evolution of the bias for which we need a model. However, as shown in Figure 11 of~\cite{Bonvin:2013ogt} this contribution is much smaller than the relativistic, wide-angle and large-angle dipoles. Therefore we can safely neglect it. Figure~\ref{fig:predictions} shows that the large-angle contribution (measured with $\beta$) is significantly larger than the other contributions.

\section{Measurements of the dipole}
\label{sec:measure}

\begin{figure}
\centering
\includegraphics[width=0.5\textwidth]{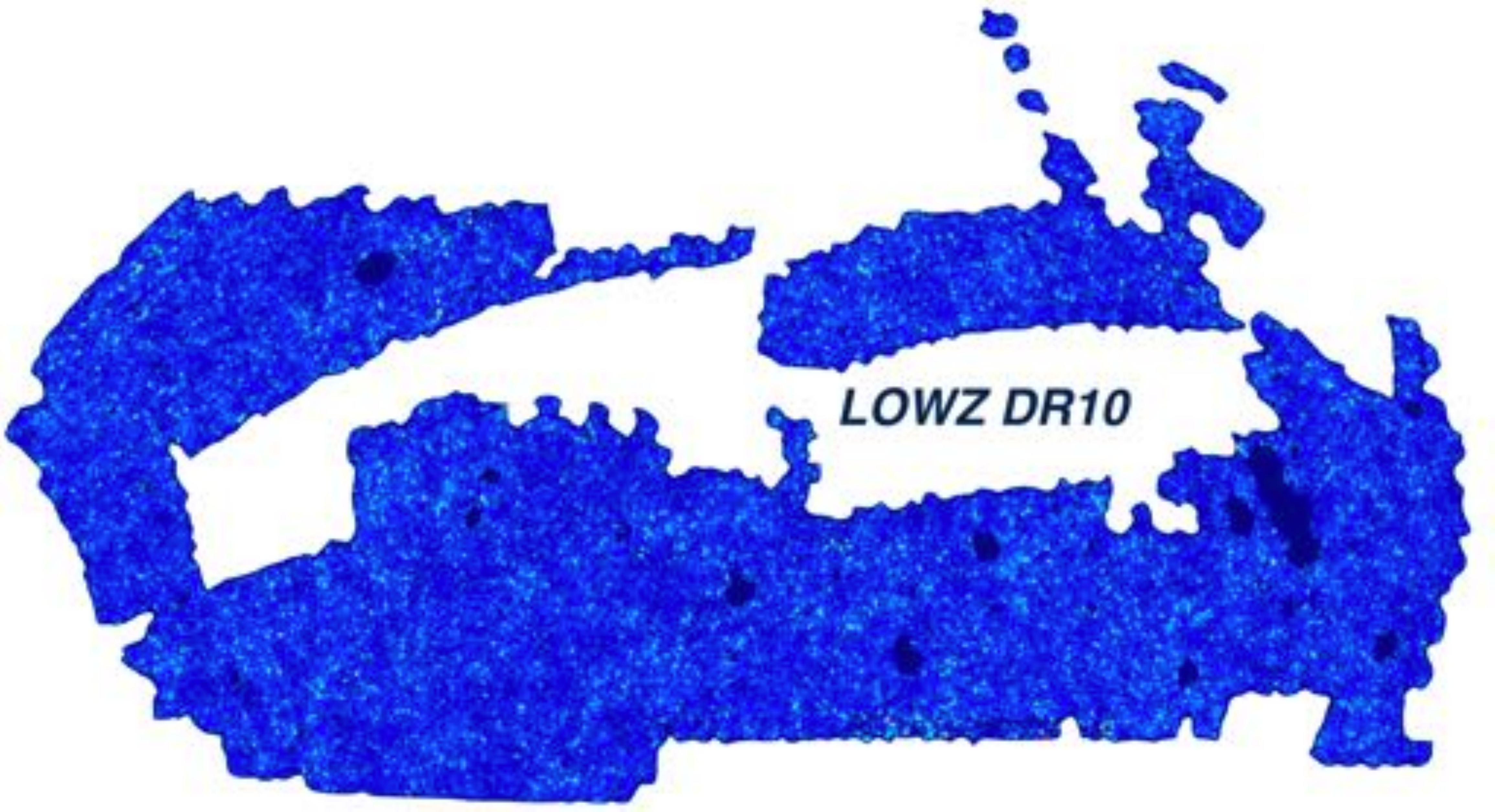}\includegraphics[width=0.5\textwidth]{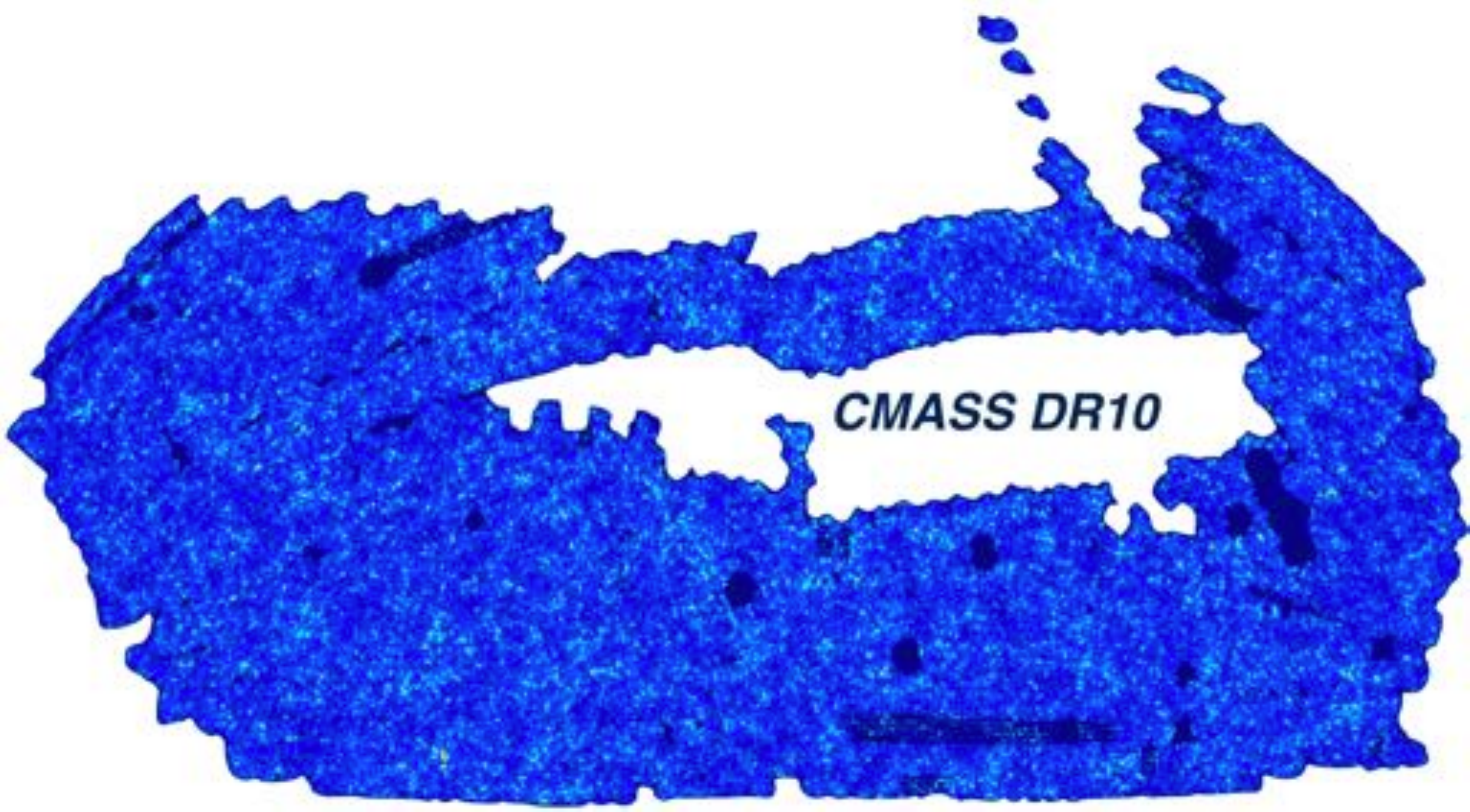}
\caption{\label{fig:mask} Distribution of galaxies in the LOWz and the CMASS samples. Weights are included.}
\end{figure}

\begin{figure}
\centering
\hspace{-1.9cm}\includegraphics[width=0.4\textwidth]{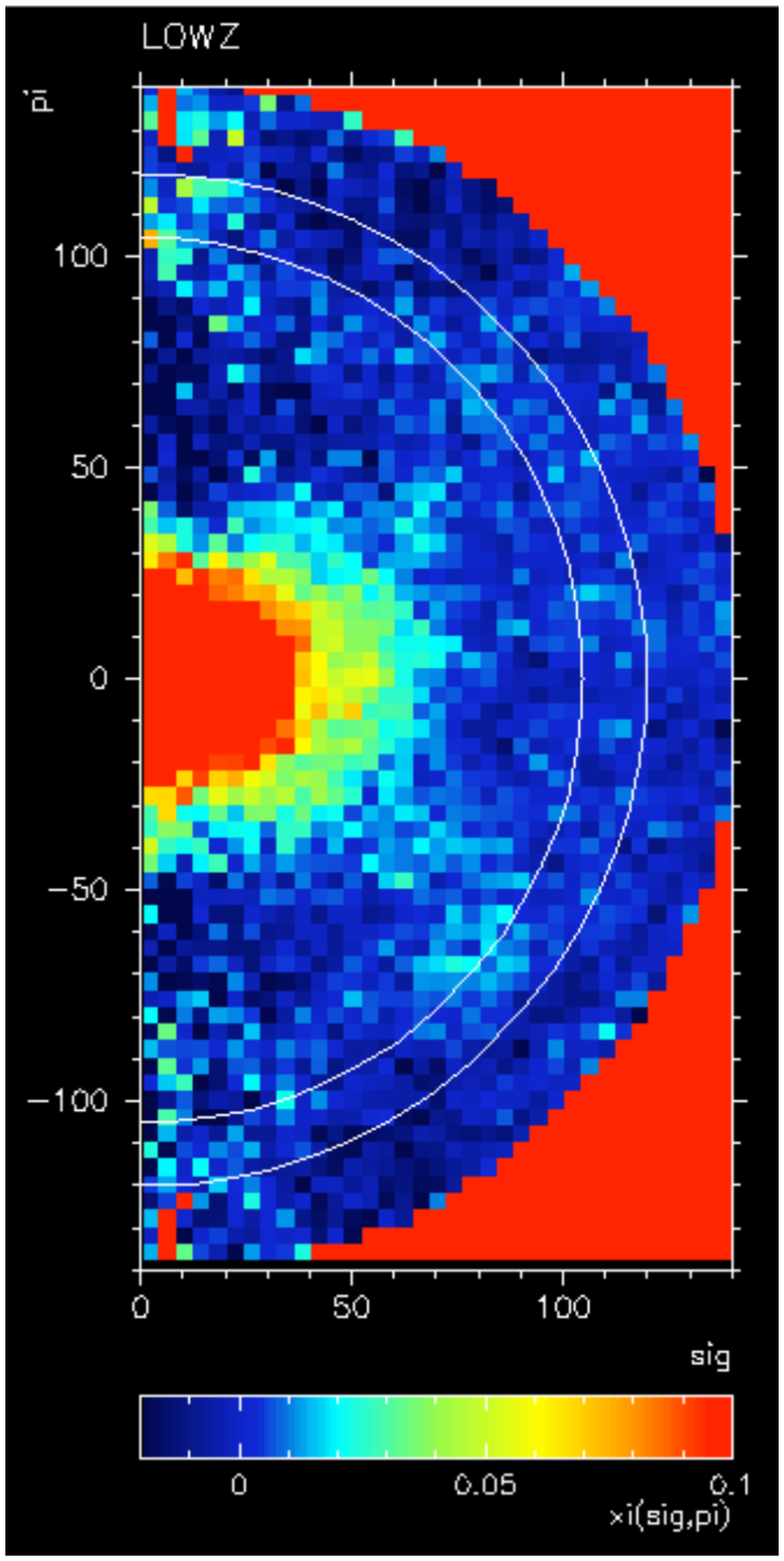}\hspace{-3cm}\includegraphics[width=0.402\textwidth]{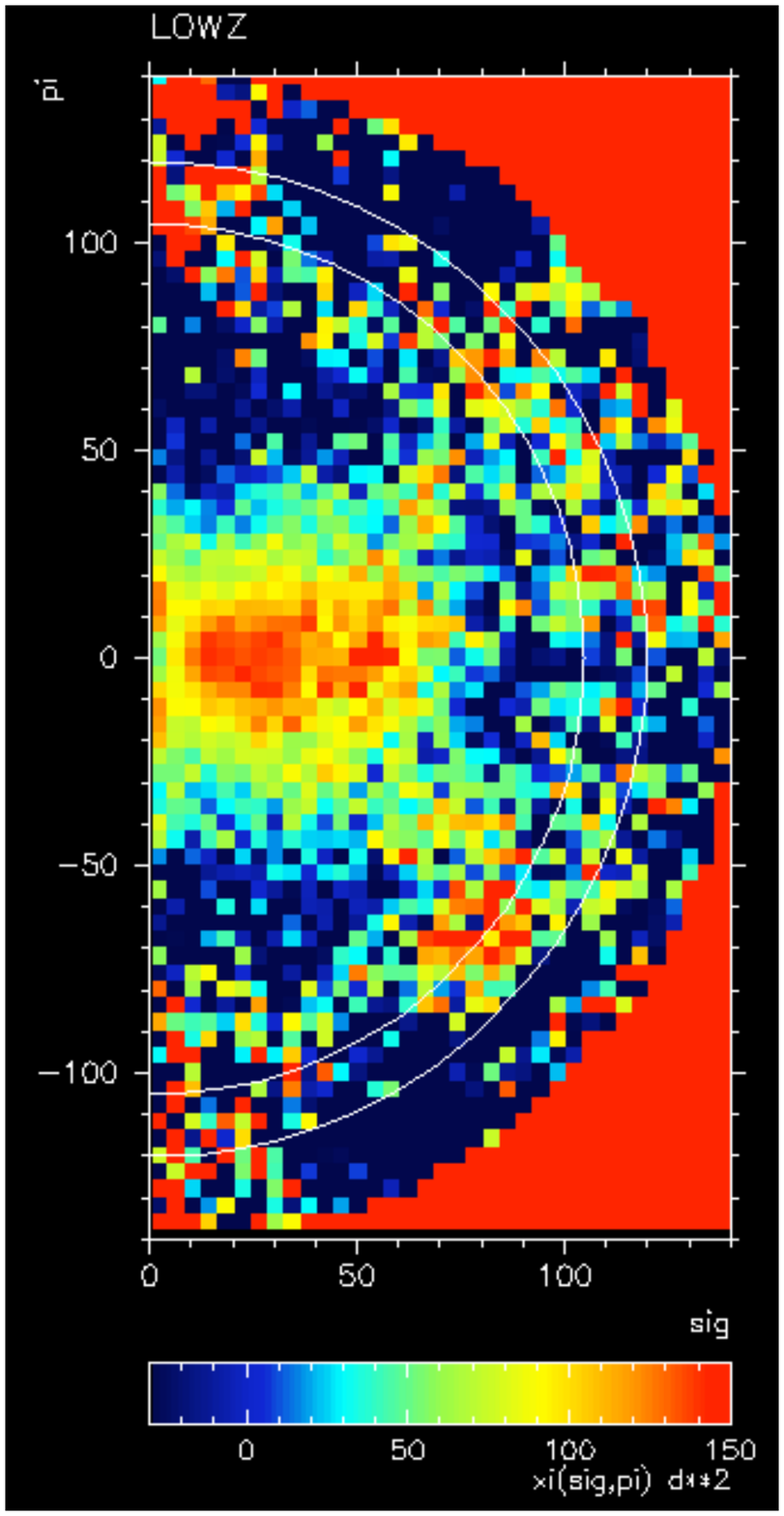}
\hspace{-3.06cm}\includegraphics[width=0.4\textwidth]{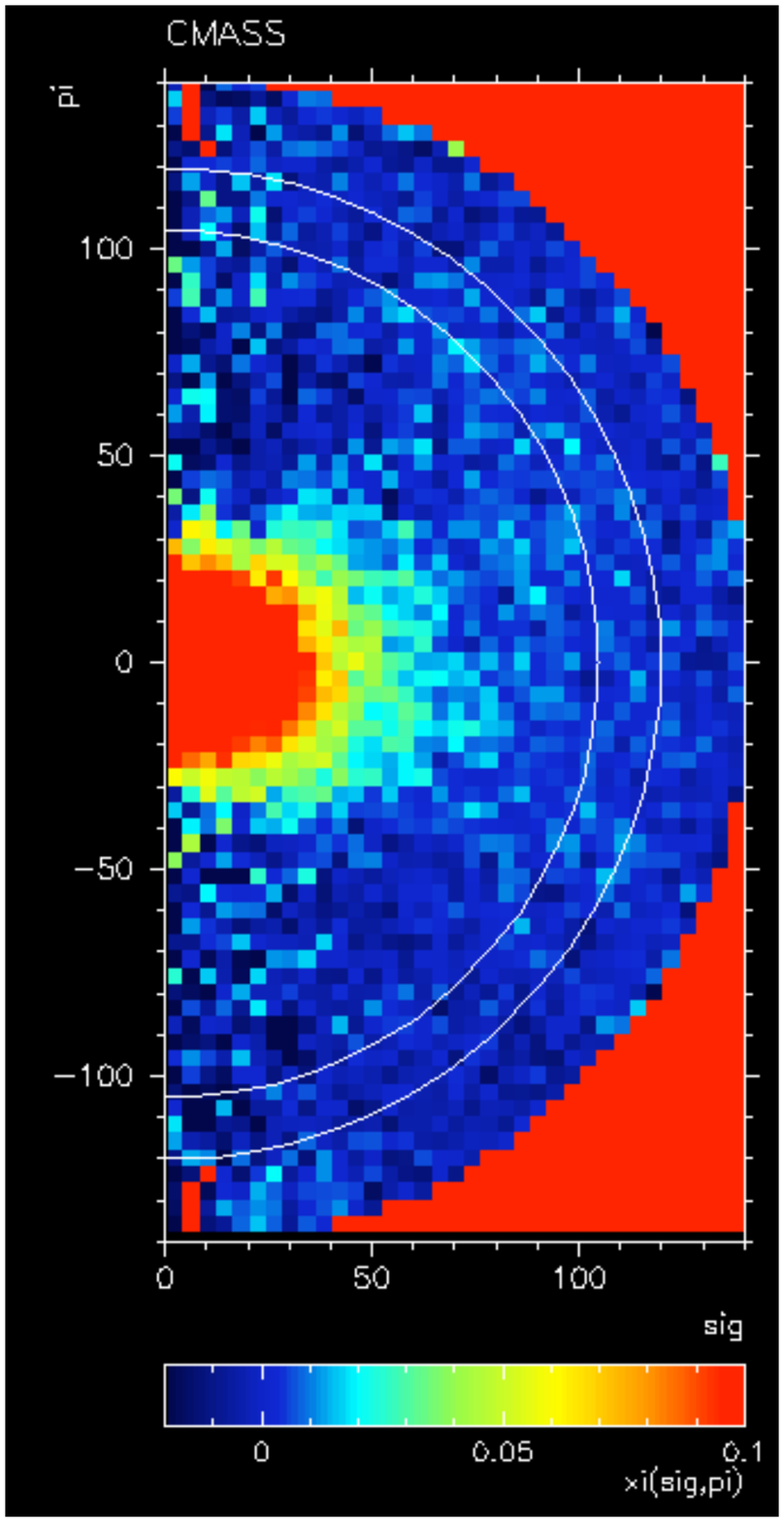}\hspace{-2.97cm}\includegraphics[width=0.4\textwidth]{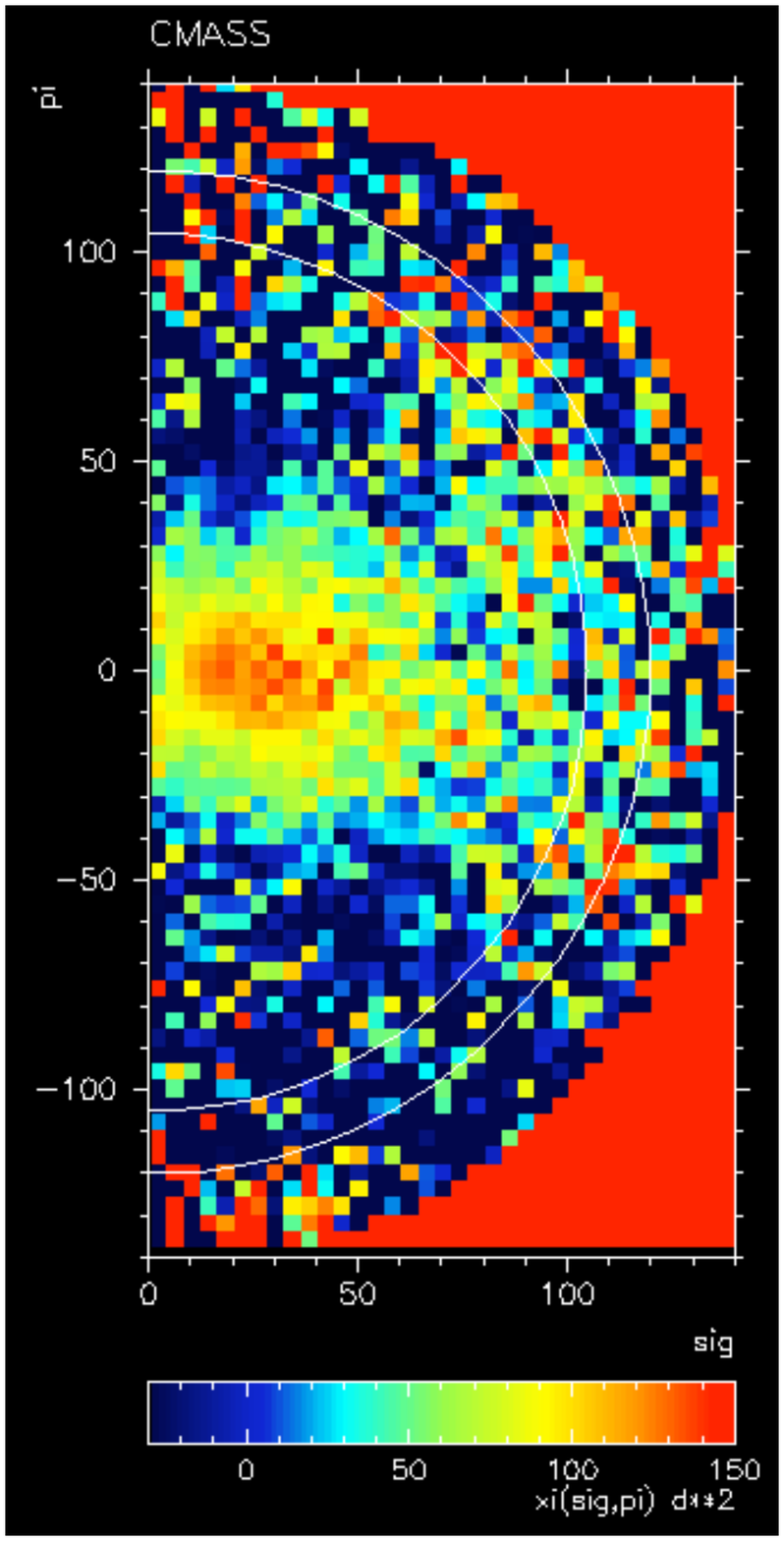}\hspace{-2cm}
\vspace{-0.5cm}\caption{\label{fig:sigmapi} 2D cross-correlation function using the angle $\sigma_{ij}$, plotted as a function of parallel (y-axis) and perpendicular (x-axis) separation with respect to the line-of-sight. The two left panels show the cross-correlation function in the LOWz sample and the two right panels in the CMASS sample. In the first and third panels we directly plot the cross-correlation function, whereas in second and fourth panels we plot the cross-correlation function multiplied by $d^2$.}
\end{figure}

\begin{figure}
\centering
\includegraphics[width=0.47\textwidth]{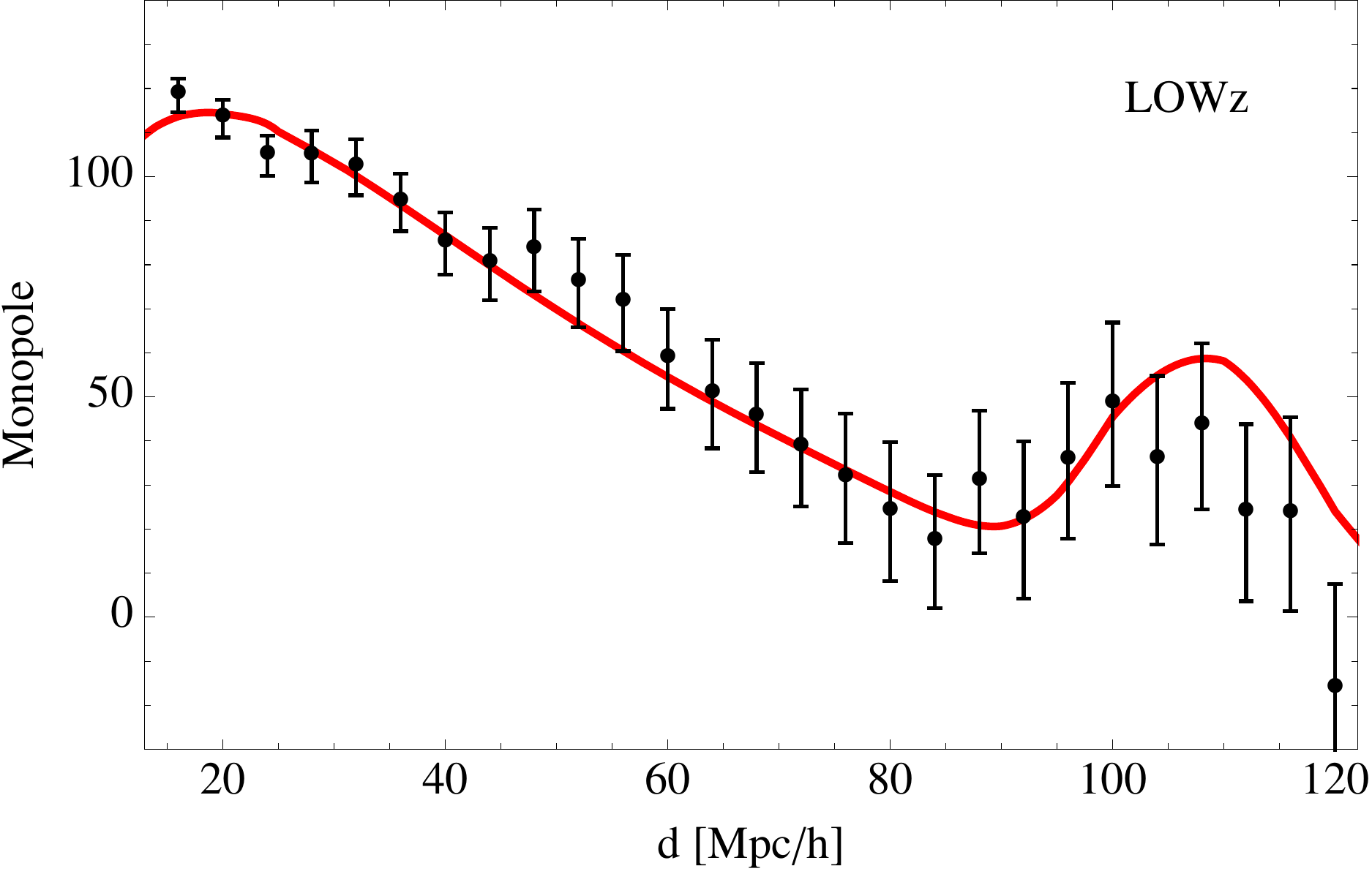}\hspace{0.8cm}
\includegraphics[width=0.47\textwidth]{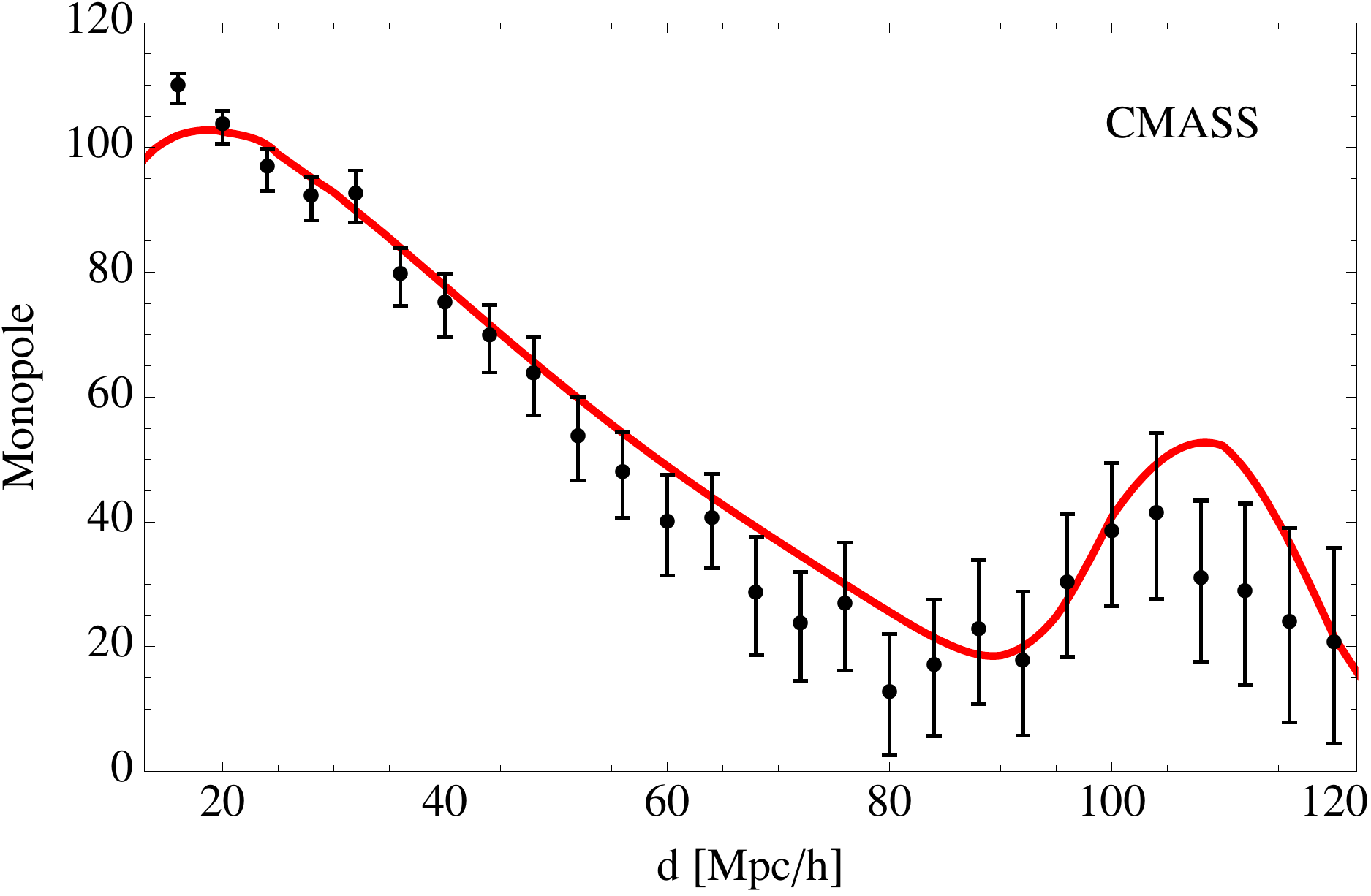}
\caption{\label{fig:BOSS_monquad} Measurements of the monopole (multiplied by $d^2$) from the cross-correlation of bright and faint galaxies in the LOWz and CMASS samples, plotted as a function of separation. The red solid line is the prediction using linear perturbation theory and the biases $b_\B$ and $b_\F$ are fitted from the total and cross monopoles, Eqs.~\eqref{monoT} and~\eqref{mono}.}
\end{figure}

We now apply our method to the BOSS survey. We measure the monopole and the dipole in the LOWz and CMASS samples, data release DR10~\cite{Anderson:2013zyy}. In each sample, we split the population of galaxies into a bright and a faint population. In order to obtain a homogeneous fraction of bright and faint galaxies over the samples, we divide them into sub-samples of width $\Delta z\simeq 0.01$ and we adjust the luminosity cut in each sub-sample to have roughly equal numbers of bright and faint galaxies. With this procedure, we obtain for both samples $\bar n_\B=0.47$ and $\bar n_\F=0.53$. In LOWz we use 148'833 galaxies and in CMASS 380'899 galaxies.
In both samples we use cubic pixels with size $\ell_p=4$\,Mpc/$h$ for separations $16\leq d\leq 120$\,Mpc/$h$. The effective redshift of the LOWz sample is $\bar z=0.303$ and of the CMASS sample $\bar z=0.575$.

We weight galaxies according to the prescriptions given in~\cite{Anderson:2013zyy} and we use the public LSS random catalogs to build the masks. In particular we use the systematics  weighting in Eq. (18) of~\cite{Anderson:2013zyy} to account for density variations due to stellar and seeing variations, redshift failures and close pairs. We also use the same FKP weighting~\cite{Feldman:1993ky}  as in~\cite{Anderson:2013zyy}. This allows us to reproduce well the monopole and quadrupole measurements shown in the top panel of Figure 18 of~\cite{Anderson:2013zyy}, including their error bars. We have checked that these results are indeed robust when we do small variations in the way we build the mask or weight the galaxies. When splitting the sample into faint and bright galaxies we use a different FKP weighting according to their relative densities, but we use the same systematics weighting  for faint and bright galaxies. This is probably a rough approximation because the faint and bright galaxies will have different sensitivities to these systematic effects. These differences will tend to cancel when we use cross-correlations between faint and bright galaxies but we have found that they are more important for the auto-correlations (faint x faint or bright x bright),  which are therefore not used in our analysis (see below).

The distribution of galaxies is shown in Figure~\ref{fig:mask} for both the LOWz and the CMASS samples.  We pixelize the samples in 3D cubical pixels and estimate the correlation function, as explained around Eq.~\eqref{estimator}. We bin the correlation in both separation and line-of-sight angle, using the angle $\sigma_{ij}$ and $\beta_{ij}$. We use 329 Jack-knife (JK) regions in LOWz and 335 JK regions in CMASS to evaluate the error-bars in the data~\footnote{Note that we have checked that the Jack-knife errors for the monopole and the quadrupole agree well with the errors from the BOSS collaboration~\cite{Anderson:2013zyy}, obtained from simulations. This gives us confidence that the Jack-knife errors on the dipole are reliable.}. The resulting 2D cross-correlations are shown in Figure~\ref{fig:sigmapi} for the angle $\sigma_{ij}$, as a function of parallel and perpendicular separation with respect to the line-of-sight (see~\cite{Gaztanaga:2008xz}). Note that contrary to the plots shown in~\cite{Gaztanaga:2008xz},  the plots in Figure~\ref{fig:sigmapi} are not symmetric in the parallel separation because we always use the bright galaxy as the centre to measure the angle and distinguish between positive and negative line-of-sight separation. This allows us to measure the multipoles as projection over Legendre polynomials. Also note in the figure some significant large scale "noise" on scales 50-100\,Mpc/$h$. This is partially due to the fact that we are measuring cross-correlations of bright and faint galaxies. These galaxies do not have the same redshift distribution or trace matter in exactly the same way, which creates some additional noise and modulation compared to the auto-correlations, particularly at large separations  (this could be called relative bias "stochasticity").
Also, note that especially for the LOWz sample, the pairs at large negative radial separation are sampling smaller volumes than the ones at positive values. This results in an enhancement of sampling variance around BAO position, as shown by the broad structure at separations (-70 radial , +70 perpendicular) in the LOWz panels~\footnote{This geometrical effect could help explaining the results in~\cite{Cabre:2010bc}, which shows how the BAO position could be well measured even at instances where sampling variance does not allow to favor a model with BAO peak from one without it.}. 

We measure the monopole using all pairs of galaxies $\zeta^{\rm T}_0(d)$ (i.e. without distinguishing bright and faint galaxies) and then using only the cross-correlation between bright and faint galaxies $\zeta_0(d)$. 
We compare these measurements with linear predictions using the fiducial cosmological parameters of~\cite{Anderson:2013zyy} ($\Omega_m=0.274$, $h=0.7$, $\Omega_b h^2=0.0224$, $n_s=0.95$ and $\sigma_8=0.8$) to find the respective bias of the two populations
\bea
\zeta^{\rm T}_0(d)&=&\left[b_{\rm T}^2+2b_{\rm T}\frac{f}{3}+\frac{f^2}{5} \right]C_0(d)\label{monoT}\, ,\\
\zeta_0(d)&=&\left[b_\B b_\F+(b_\B+b_\F)\frac{f}{3}+\frac{f^2}{5} \right]C_0(d)\label{mono}\, ,
\eea 
where the bias of the whole sample $b_{\rm T}$ is related to the bias of the bright and faint populations by
\be
b_{\rm T}=\bar n_\B b_\B+\bar n_\F b_\F\, .
\ee
In LOWz we find $b_\B=2.30\pm 0.21$ and $b_\F=1.31\pm 0.21$; and in CMASS we find $b_\B=2.36\pm 0.18$ and $b_\F=1.46\pm 0.18$. We then use these values to predict the amplitude of the dipole. Note that this method is more robust than fitting the bias using separately the monopole from the bright $\zeta^\B_0(d)$ and the monopole from the faint $\zeta^\F_0(d)$. The reason is that the weights defined in~\cite{Anderson:2013zyy} are constructed for the whole population. Applying these weights separately to the bright and faint populations generates inhomogeneous samples. This significantly affects the measurement of the bright and faint monopoles. The measurement of the cross-correlation between the bright and the faint populations is however relatively insensitive to this problem and it provides therefore a more robust way of extracting the biases.

In Figure~\ref{fig:BOSS_monquad} we show the measurements of the monopole for the cross-correlations between the bright and the faint populations in the LOWz and CMASS samples. The red line represents the theoretical prediction.

\subsection{Measurement of the dipole using the median angle $\mu_{12}\equiv\cos\sigma$}

\begin{figure}
\centering
\includegraphics[width=0.47\textwidth]{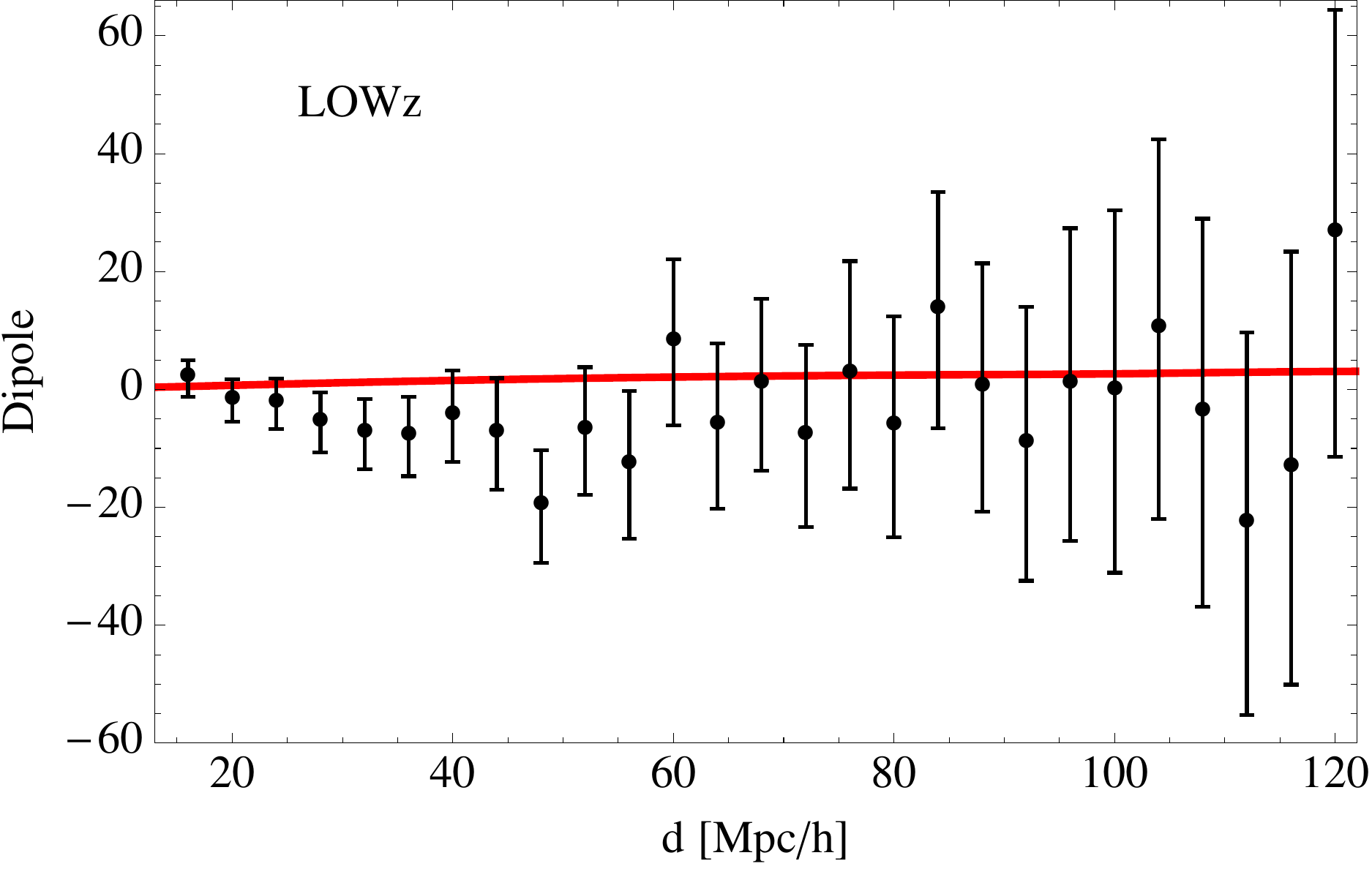}\hspace{0.8cm}\includegraphics[width=0.47\textwidth]{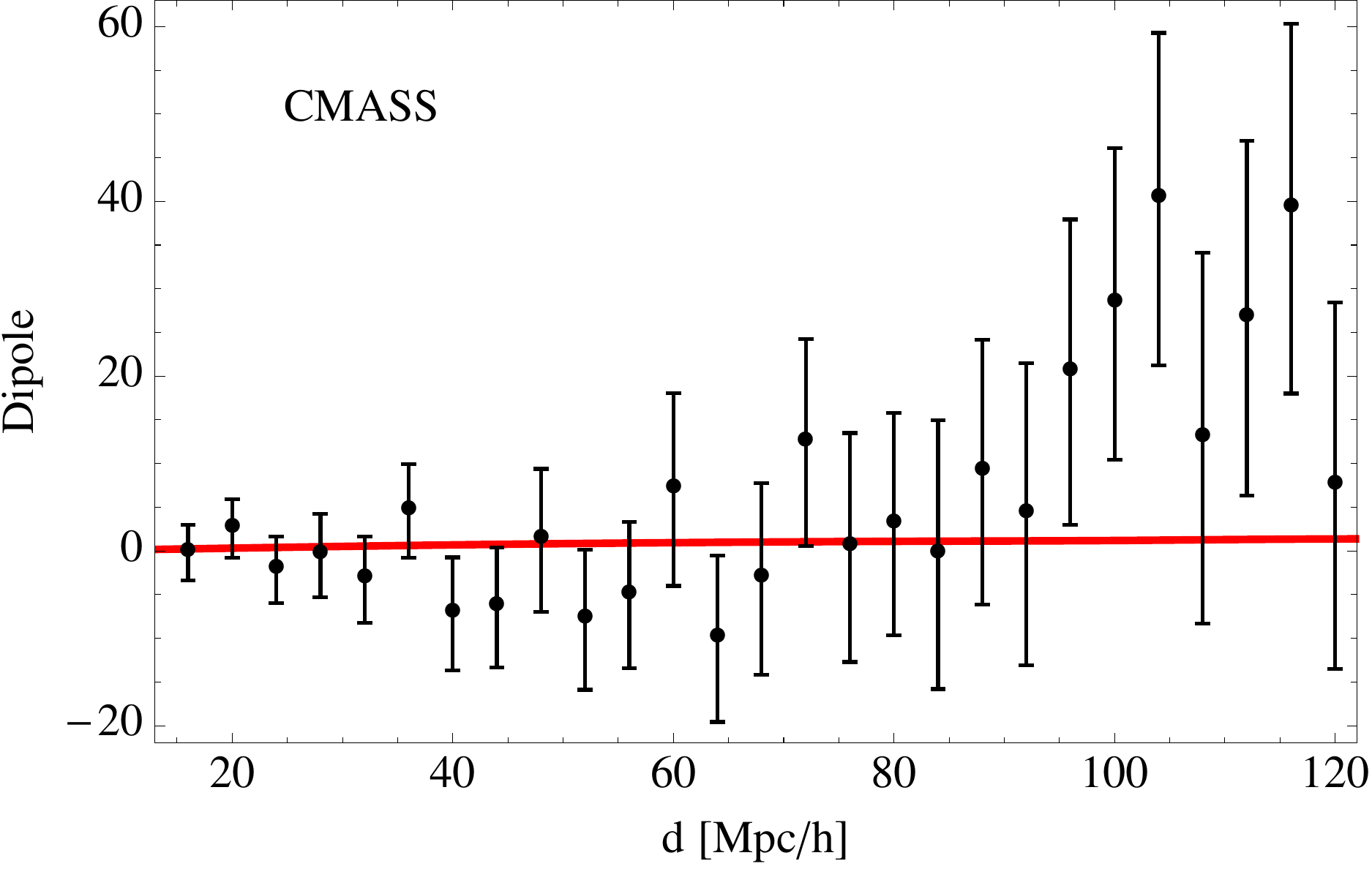}
\caption{\label{fig:dipole1} Measurements of the dipole using the median angle $\mu_{12}=\cos\sigma$, multiplied by $d^2$, from the cross-correlation of bright and faint galaxies in the LOWz and CMASS samples, plotted as a function of separation. The red solid line is the prediction using linear perturbation theory and the biases $b_\B$ and $b_\F$ are fitted from the total and cross monopoles, Eqs.~\eqref{monoT} and~\eqref{mono}.}
\end{figure}

\begin{figure}
\centering
\includegraphics[width=0.47\textwidth]{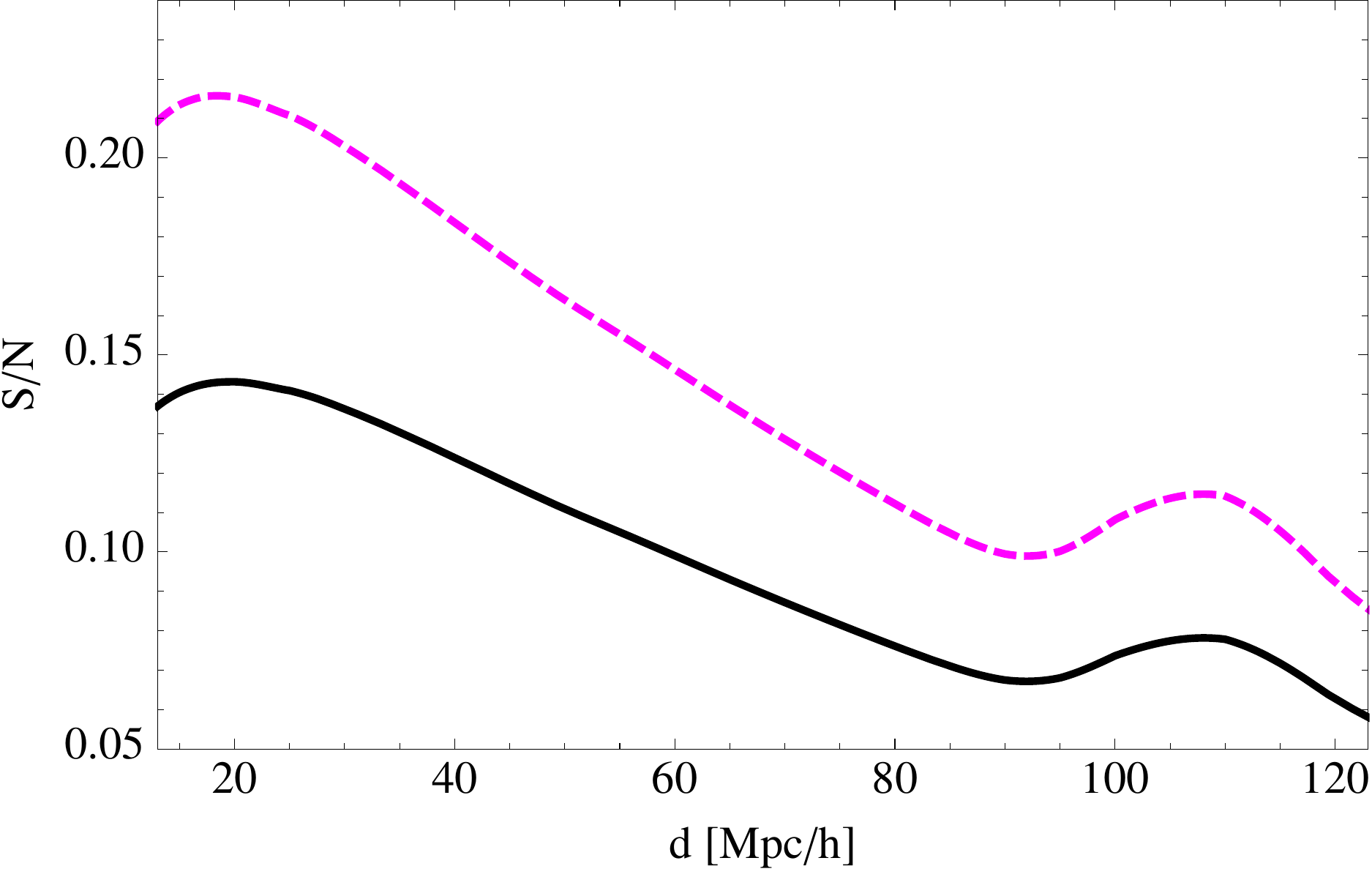}
\caption{\label{fig:SN1} Signal-to-noise for the dipole using the median angle $\mu_{12}=\cos\sigma$, in the LOWz sample (magenta dashed line) and in the CMASS sample (black solid line), plotted as a function of separation.}
\end{figure}

We start by measuring the dipole with the median angle $\mu_{12}=\cos\sigma$. In Figure~\ref{fig:dipole1} we show the dipole in the LOWz and CMASS samples. In both cases, the dipole is compatible with zero within the error bars. 

In~\cite{Bonvin:2015kuc}, we have calculated the signal-to-noise of the dipole. We found that the signal-to-noise at fixed separation $d$ is given by
\begin{align}
\label{SNfin}
\frac{S}{N}(d)=\sqrt{\frac{2\pi N_{\rm tot}}{3}}
\frac{2\bar n_\B\bar n_\F(b_\B-b_\F)\,\alpha(d)}
{\left[2\bar n_\B\bar n_\F\left(\bar N d^2 \ell_p \right)^{-1}
+3 \Big(D_0\sigma_0(d, d)+D_2\sigma_2(d, d)+D_4\sigma_4(d, d)\Big)\right]^{1/2}}\, ,
\end{align}
where 
\be
\alpha(d)=\frac{1}{2\pi^2}\left(\frac{\dot\HH}{\HH^2}+\frac{2}{r\HH} \right)\frac{\HH}{\HH_0}f
\int dk \,k \HH_0P(k, \bar z)j_1(k d)\, .
\ee
The functions $\sigma_\ell$ are given by 
\begin{align}
\sigma_\ell(d, d')=-\frac{1}{2\pi^2}\int_{-1}^1 d\rho\, \rho\int_{-1}^1 d\nu\, \nu\int_0^{2\pi}d\varphi\int dk k^2 P(k, \bar z) j_\ell(ks)P_\ell\left(\frac{d\rho+d'\nu}{s}\right)\, ,\hspace{0.3cm}\ell=0,2,4\,
\end{align}
with $P_\ell$ the Legendre polynomial of degree $\ell$ and
\be
s=\sqrt{d^2+d'^2+2dd'\big(\rho\nu+\sqrt{(1-\rho^2)(1-\nu^2)}\sin\varphi \big)}\, .
\ee 
For two populations of galaxies we have
\bea
D_0&=&\bar n^2_\B \bar n_\F\left(b_\B^2+\frac{2b_\B f}{3}+\frac{f^2}{5} \right)
+\bar n^2_\F \bar n_\B\left(b_\F^2+\frac{2b_\F f}{3}+\frac{f^2}{5} \right)\, ,\\
D_2&=&-\bar n^2_\B \bar n_\F\left(\frac{4b_\B f}{3}+\frac{4f^2}{7} \right)-\bar n^2_\F \bar n_\B\left(\frac{4b_\F f}{3}+\frac{4f^2}{7} \right)\, ,\\
D_4&=&\big(\bar n^2_\B \bar n_\F+\bar n^2_\F \bar n_\B\big)\frac{8f^2}{35}\, .
\eea

Using these expressions we can calculate the signal-to-noise for the dipole in the LOWz and in the CMASS samples. Since the number densities and the fraction of bright and faint galaxies evolve with redshift, we split each of the samples into 3 redshift bins. We calculate the signal-to-noise in each bin and then add them in quadrature to obtain the total signal-to-noise (note that to a very good approximation the different redshift bins are uncorrelated). We plot the results in Figure~\ref{fig:SN1}. The signal-to-noise is very small showing that no detection of the dipole is possible using the median angle. Note that the signal-to-noise calculated here is the signal-to-noise at fixed separation $d$. The cumulative signal-to-noise combining all separations can be calculated using Eq.~(55) of~\cite{Bonvin:2015kuc}.

\subsection{Measurement of the dipole using the angle from the bright galaxy $\mu_1\equiv \cos\beta$}

\begin{figure}
\centering
\includegraphics[width=0.47\textwidth]{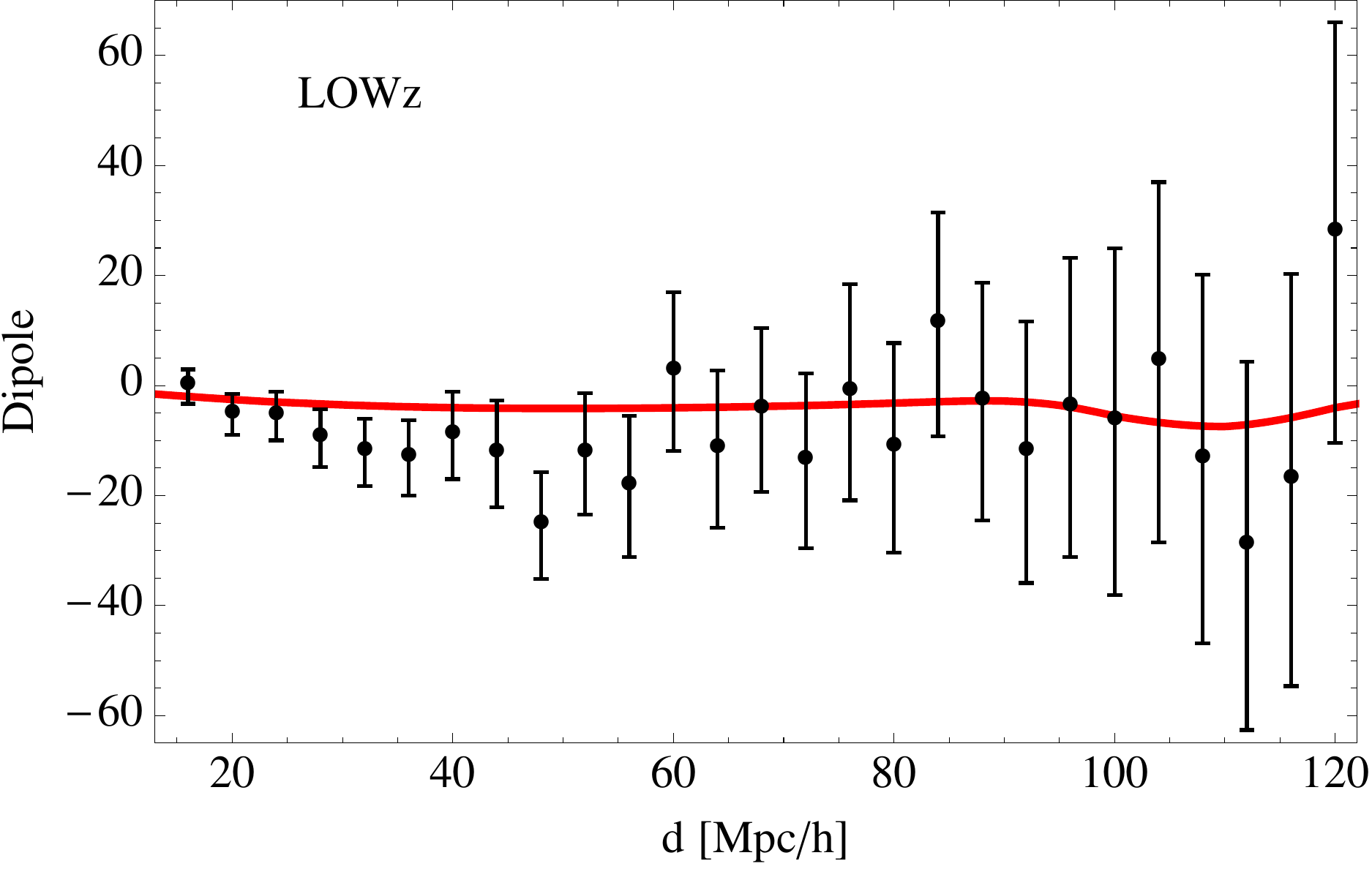}\hspace{0.8cm}\includegraphics[width=0.47\textwidth]{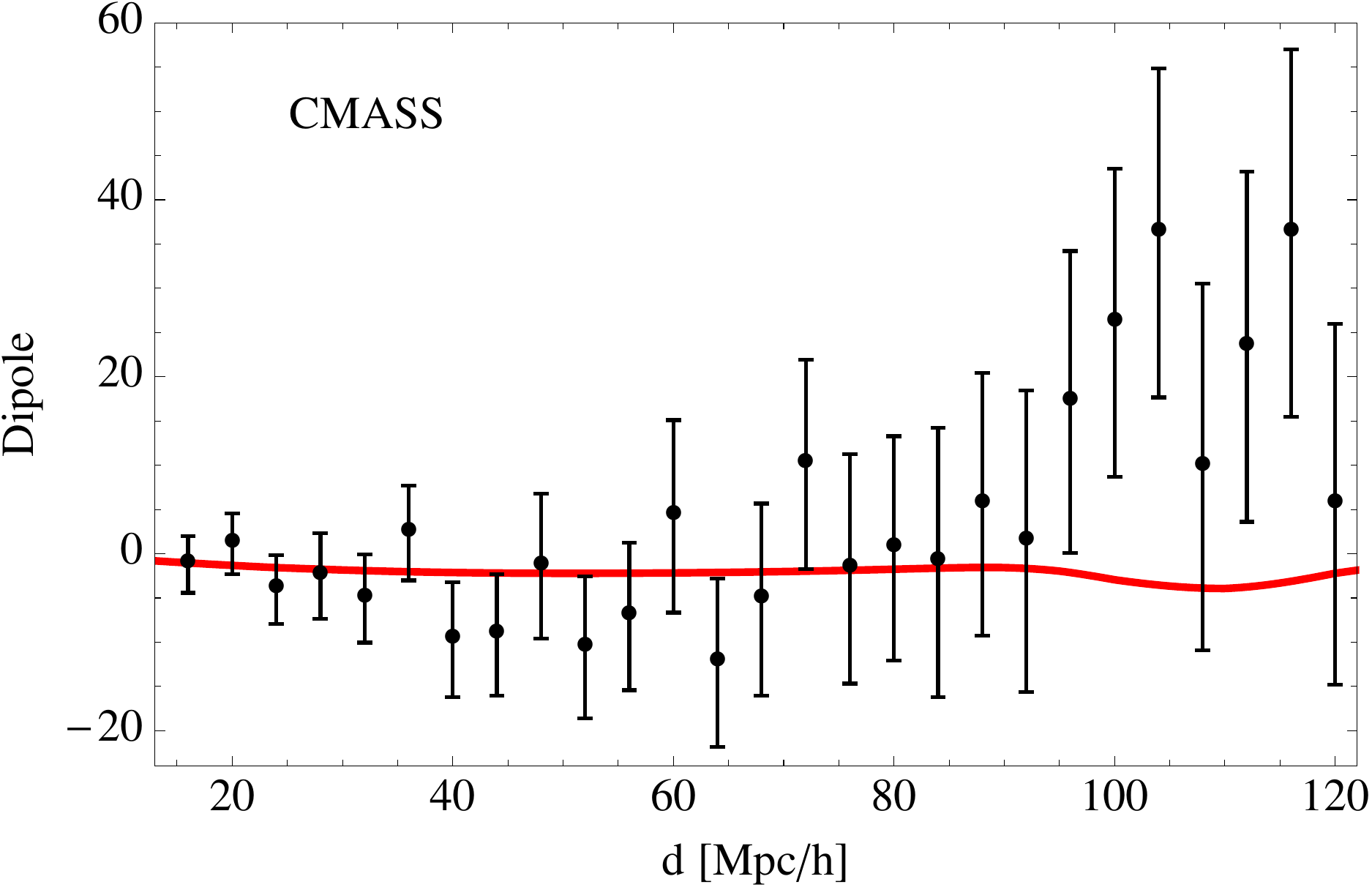}
\caption{\label{fig:dipole2} Measurements of the dipole using the angle from the bright galaxy $\mu_1= \cos\beta$, multiplied by $d^2$, from the cross-correlation of bright and faint galaxies in the LOWz and CMASS samples, plotted as a function of separation. The red solid line is the prediction using linear perturbation theory and the biases $b_\B$ and $b_\F$ are fitted from the total and cross monopoles, Eqs.~\eqref{monoT} and~\eqref{mono}.}
\end{figure}

\begin{figure}
\centering
\includegraphics[width=0.47\textwidth]{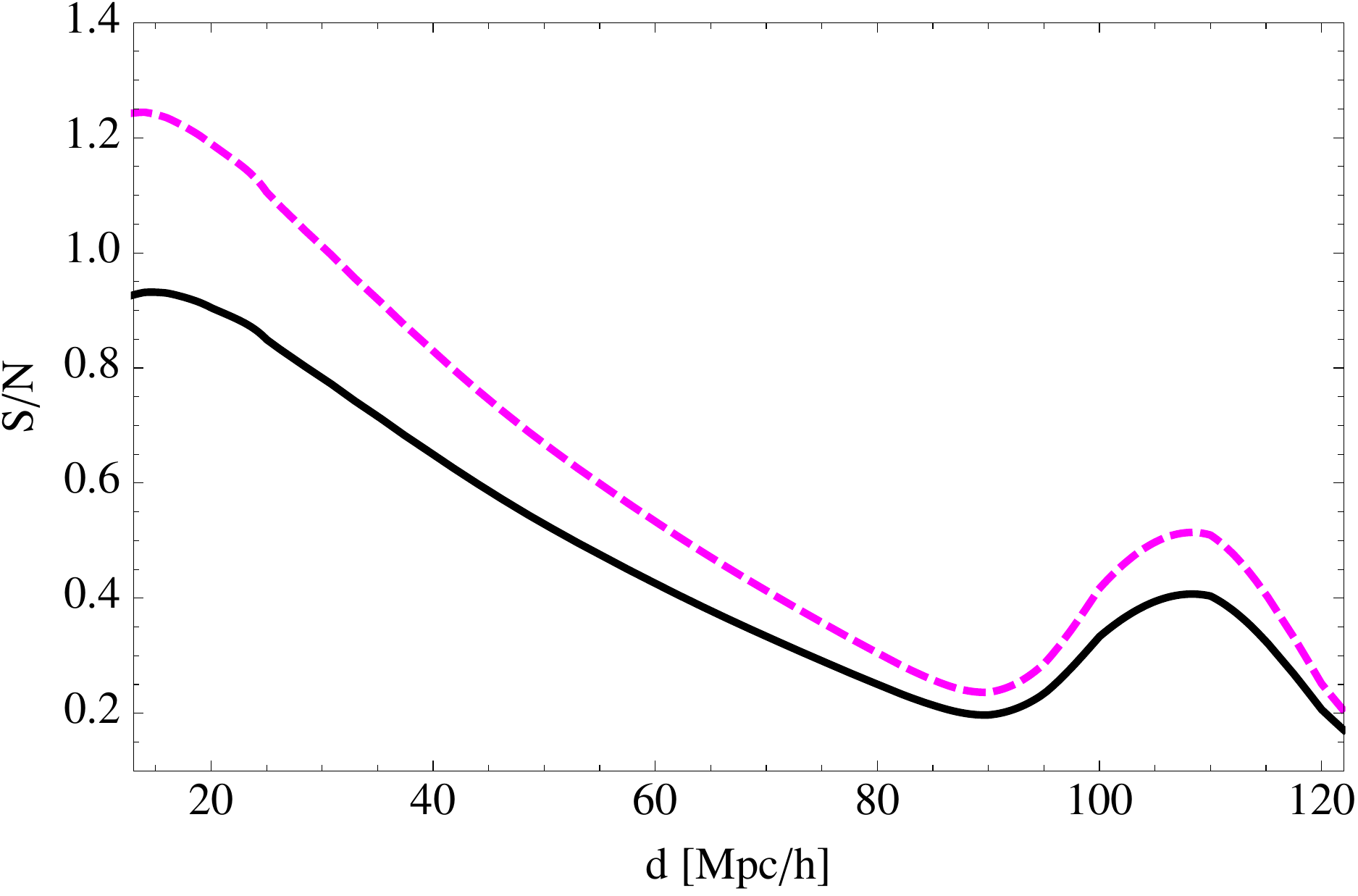}
\caption{\label{fig:SN2} Signal-to-noise for the dipole using the angle from the bright galaxy $\mu_1= \cos\beta$, in the LOWz sample (magenta dashed line) and in the CMASS sample (black solid line), plotted as a function of separation.}
\end{figure}

We then measure the dipole using the angle from the bright galaxy $\mu_1\equiv \cos\beta$. As shown in Figure~\ref{fig:predictions}, with this choice of kernel the dipole is much larger that with the median angle $\sigma$, due to the existence of the large-angle dipole. In Figure~\ref{fig:dipole2} we show the dipole in the LOWz and CMASS samples. Even though the signal is much larger in this case, the errors are still too large to allow for a detection of the large-angle dipole.
In Figure~\ref{fig:SN2} we show the predicted signal-to-noise in this case, which is of order 1 at small separation.  

\subsection{Measurement of the large-angle dipole using the angle difference $\mu_1-\mu_{12}=\cos\beta-\cos\sigma$}
\label{sec:case 3}

\begin{figure}
\centering
\includegraphics[width=0.47\textwidth]{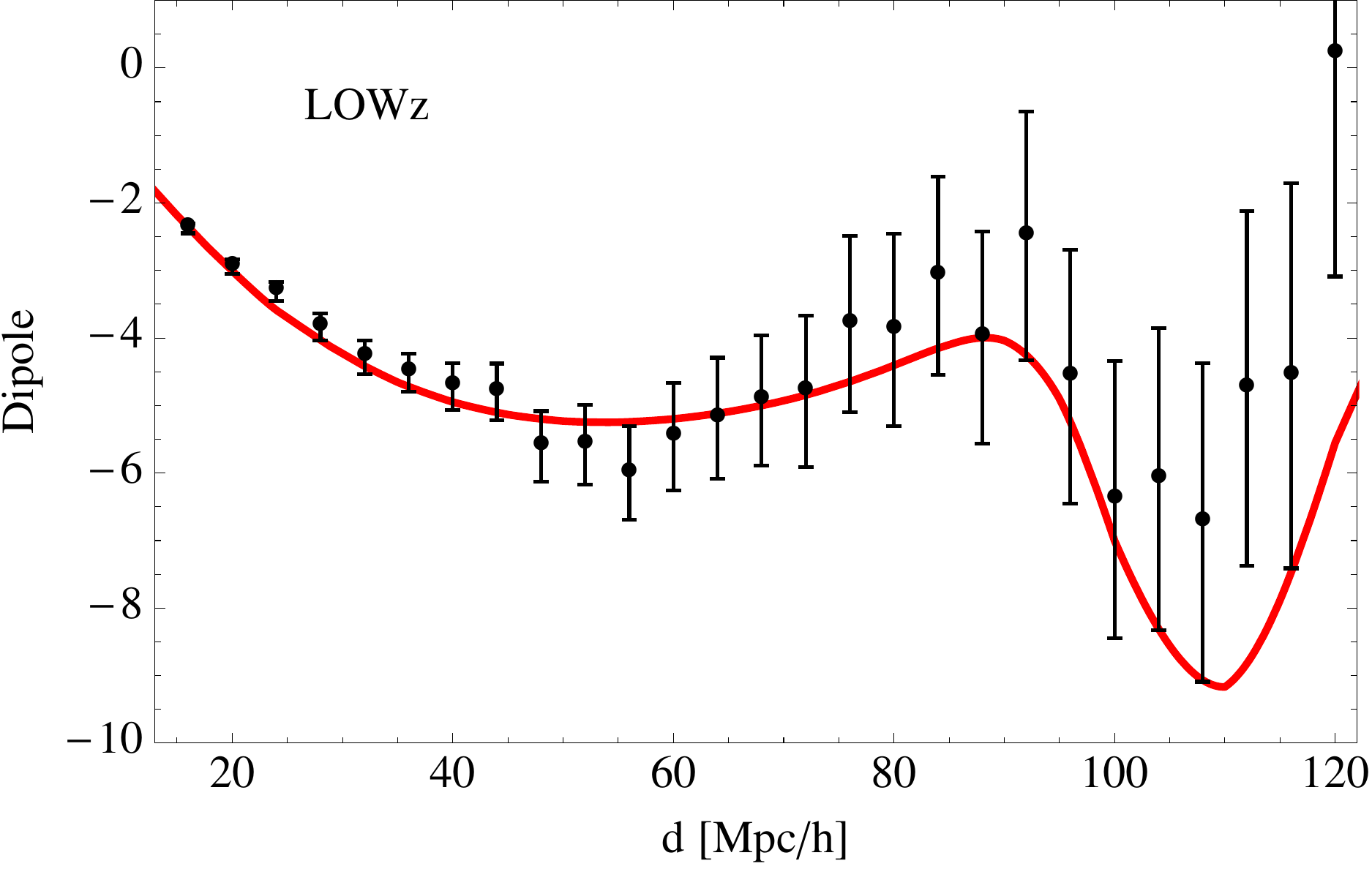}\hspace{0.8cm}\includegraphics[width=0.47\textwidth]{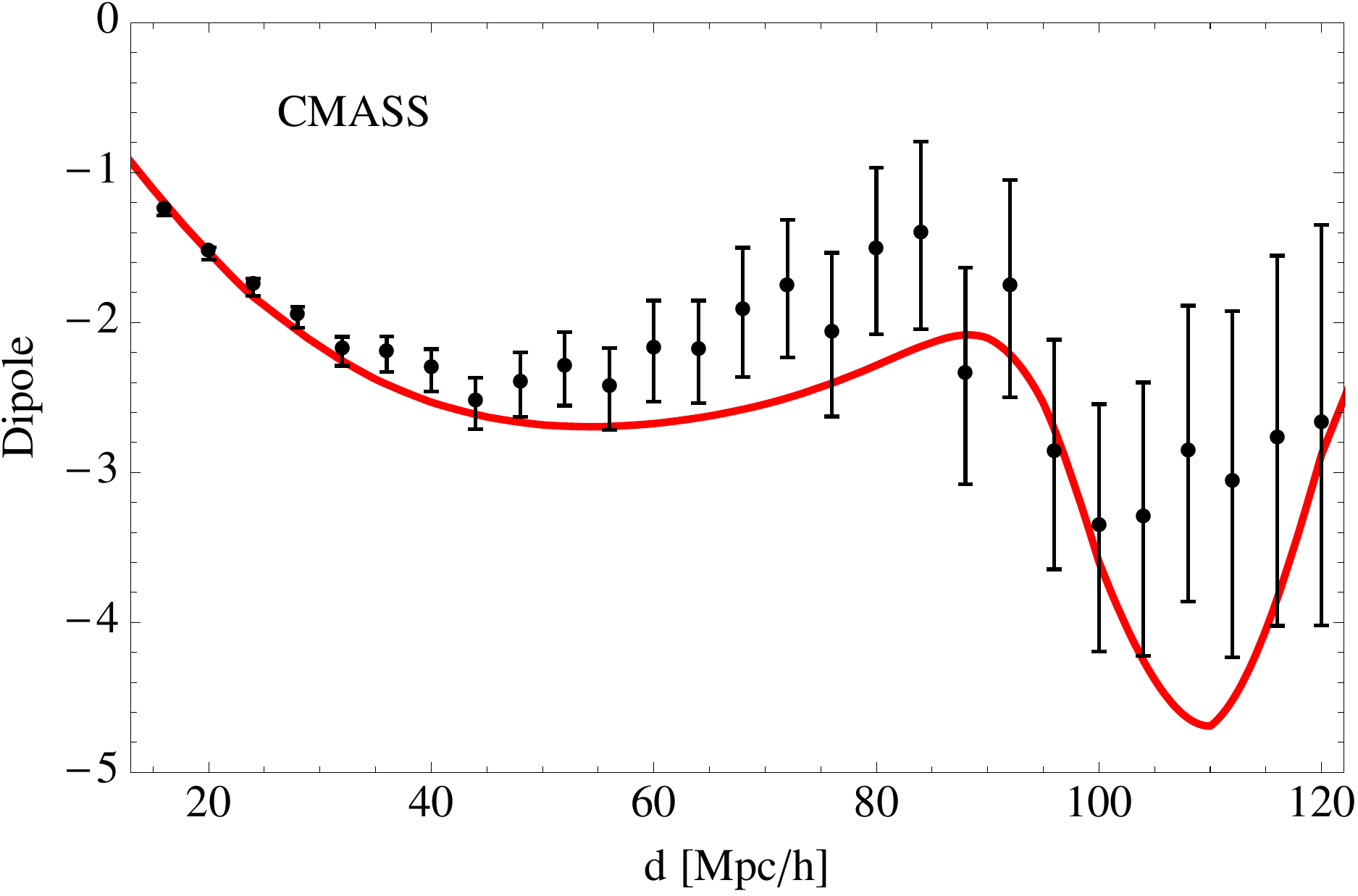}
\caption{\label{fig:dipole3} Measurements of the dipole using $\mu_1-\mu_{12}=\cos\beta-\cos\sigma$ (kernel~\eqref{kernel3}), multiplied by $d^2$, from the cross-correlation of bright and faint galaxies in the LOWz and CMASS samples, plotted as a function of separation. The red solid line is the prediction using linear perturbation theory and the biases $b_\B$ and $b_\F$ are fitted from the total and cross monopoles, Eqs.~\eqref{monoT} and~\eqref{mono}.}
\end{figure}

From Eqs.~\eqref{rel},~\eqref{evol},~\eqref{wide1} and~\eqref{wide2}, we see that we can construct a new kernel to isolate the large-angle effect. Indeed using 
\be
\label{kernel3}
\w{i}{j}=\norm \big(\cos\beta_{ijL_i L_j}-\cos\sigma_{ijL_i L_j}\big)\delta_K(d_{ij}-d)\, ,
\ee
we obtain
\begin{align}
\label{widedipole}
\langle \hat\xi \rangle&=-\norm\dn_{\B}\dn_{\F}\sum_{ij}
\langle \Delta_\B(\bx_j)\Delta_\F(\bx_i)\rangle\frac{d_{ij}}{r_i}(1\!-\!\cos^2\beta_{ij})\delta_K(d_{ij}-d)\\
&=\langle \hat\xi^{\rm large} \rangle=-\frac{d}{r}\left[\zeta_0(d)-\frac{1}{5}\zeta_2(d)\right]\, , \nonumber
\end{align}
where $\zeta_0$ is the monopole defined in Eq.~\eqref{mono} and $\zeta_2$ is the quadrupole given by
\be
\zeta_2(d)=-\left[(b_\B+b_\F)\frac{2f}{3}+\frac{4f^2}{7} \right]C_2(d)\, .\label{quad}
\ee

The relativistic dipole, the dipole due to evolution and the wide-angle dipole are therefore totally absent with this choice of kernel: only the large-angle effect contributes.
In Figure~\ref{fig:dipole3} we show the dipole measured with this new kernel in the LOWz and CMASS samples. We have a very clear detection of the large-angle dipole, which agrees well with the theoretical predictions in both samples. The signal-to-noise for the dipole reaches 50 at small separation. As seen from Eq.~\eqref{widedipole}, this large-angle dipole does not contain new statistical information since it is just a geometrical combination of the monopole and the quadrupole. However the detection of this effect and its agreement with theoretical predictions validate our method for extracting the dipole from the two-point correlation function. Moreover, the large-angle dipole could provide an alternative way of measuring redshift-space distortions with different systematics. As shown in Section~\ref{sec:1pop}, the quadrupole extracted from the dipole is indeed not exactly equivalent to the quadrupole measured directly from the two-point function, which suggests that the various multipoles are affected differently by systematics.

The combination of angles $\mu_{1}-\mu_{12}$ allows us to clearly detect the large-angle dipole. The reason is that this combination removes the dominant part of the variance. Using kernel~\eqref{kernel3} the variance of the dipole can indeed be written as
\begin{align}
{\rm var}(\hat\xi^{\rm large})=2\norm^2\sum_{ijL_i L_j}\sum_{abL_a L_b}&\delta_K(d_{ij}-d)\delta_K(d_{ab}-d)\big(\cos\beta_{ijL_i L_j}-\cos\sigma_{ijL_i L_j}\big)\big(\cos\beta_{abL_a L_b}-\cos\sigma_{abL_a L_b}\big)\nonumber\\
&\times\langle\delta n_{L_i}(\bx_i)\delta n_{ L_a}(\bx_a)\rangle
\langle\delta n_{L_j}(\bx_j)\delta n_{ L_b}(\bx_b)\rangle\, .
\end{align}
In the distant-observer approximation $\cos\beta_{ijL_i L_j}=\cos\sigma_{ijL_i L_j}$ and $\cos\beta_{abL_a L_b}=\cos\sigma_{abL_a L_b}$ so that the variance exactly vanishes. This similarity in the variance is clearly visible if we compare Figures~\ref{fig:dipole1} and~\ref{fig:dipole2}. The amplitude of the error bars as well their variation with separation look exactly the same. As a consequence with kernel~\eqref{kernel3} the only contribution to the variance is from large-angle effects, which are suppressed by a factor $d/r$. 

\subsection{Forecasts for future surveys}
\label{sec:forecasts}

Our analysis shows that the relativistic dipole is too small to be detected in the LOWz and CMASS samples. The BOSS survey turns out however not to be the optimal survey to measure the relativistic dipole. First the mean redshift of the survey is relatively high ($\bar z=0.303$ for LOWz and $\bar z=0.575$ for CMASS). The relativistic distortions decrease quickly with redshift, meaning that a survey at lower redshift may be more appropriate. Second, BOSS selects Luminous Red Galaxies (LRG), which are quite similar in terms of bias. A survey with more diverse types of galaxies would therefore be better to measure the relativistic dipole. In a companion paper~\cite{Bonvin:2015kuc}, we have calculated the signal-to-noise of the relativistic dipole in the up-comming  DESI (Dark Energy Spectroscopic Instrument) Bright Galaxy Survey~\cite{2015AAS...22533610C} (see also Eq.~\eqref{SNfin}). With two populations of galaxies we found a cumulative signal-to-noise of 5.8 over the range of separation $8\leq d\leq 120$\,Mpc/$h$ and redshift $z<0.3$. This shows that the relativistic dipole should be detectable in the near future.

Another limitation of our analysis is that we split the survey into two populations only. As a consequence all the correlations within the bright population and within the faint population are lost. A more clever splitting may increase the signal-to-noise of the dipole. In~\cite{Bonvin:2015kuc}, we showed how to construct an optimal estimator to measure the relativistic dipole with an arbitrary number of galaxies' populations. Using our optimal kernel with 6 populations of galaxies we found a cumulative signal-to-noise of 7.4 in DESI. Note that those numbers refer to the relativistic dipole only. As shown in~\cite{Bonvin:2013ogt}, the wide-angle dipole can be subtracted from the signal using the measurement of the quadrupole of each of the populations.

\begin{figure}
\centering
\includegraphics[width=0.47\textwidth]{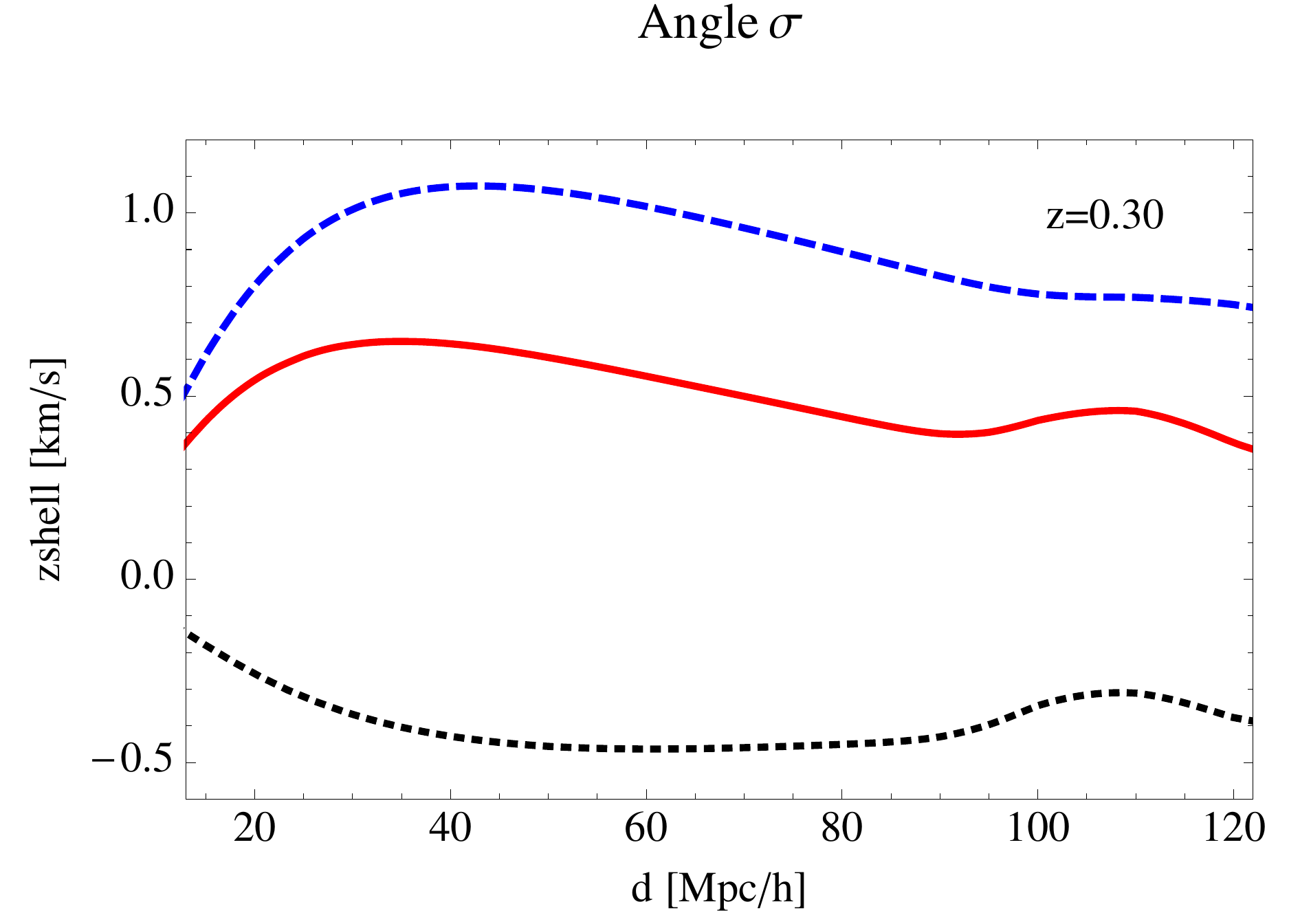}\hspace{0.8cm}\includegraphics[width=0.47\textwidth]{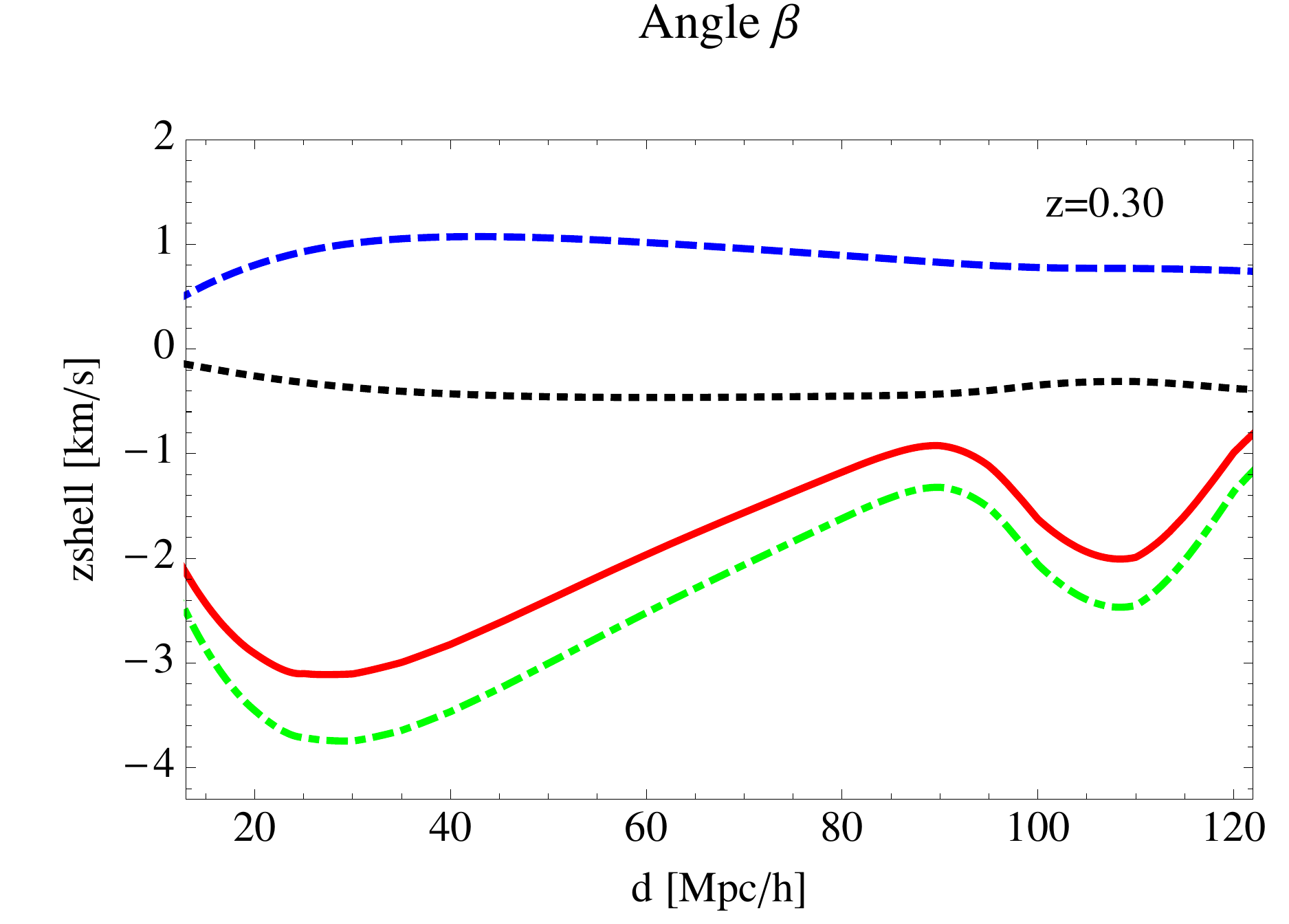}\\
\vspace{0.5cm}\includegraphics[width=0.47\textwidth]{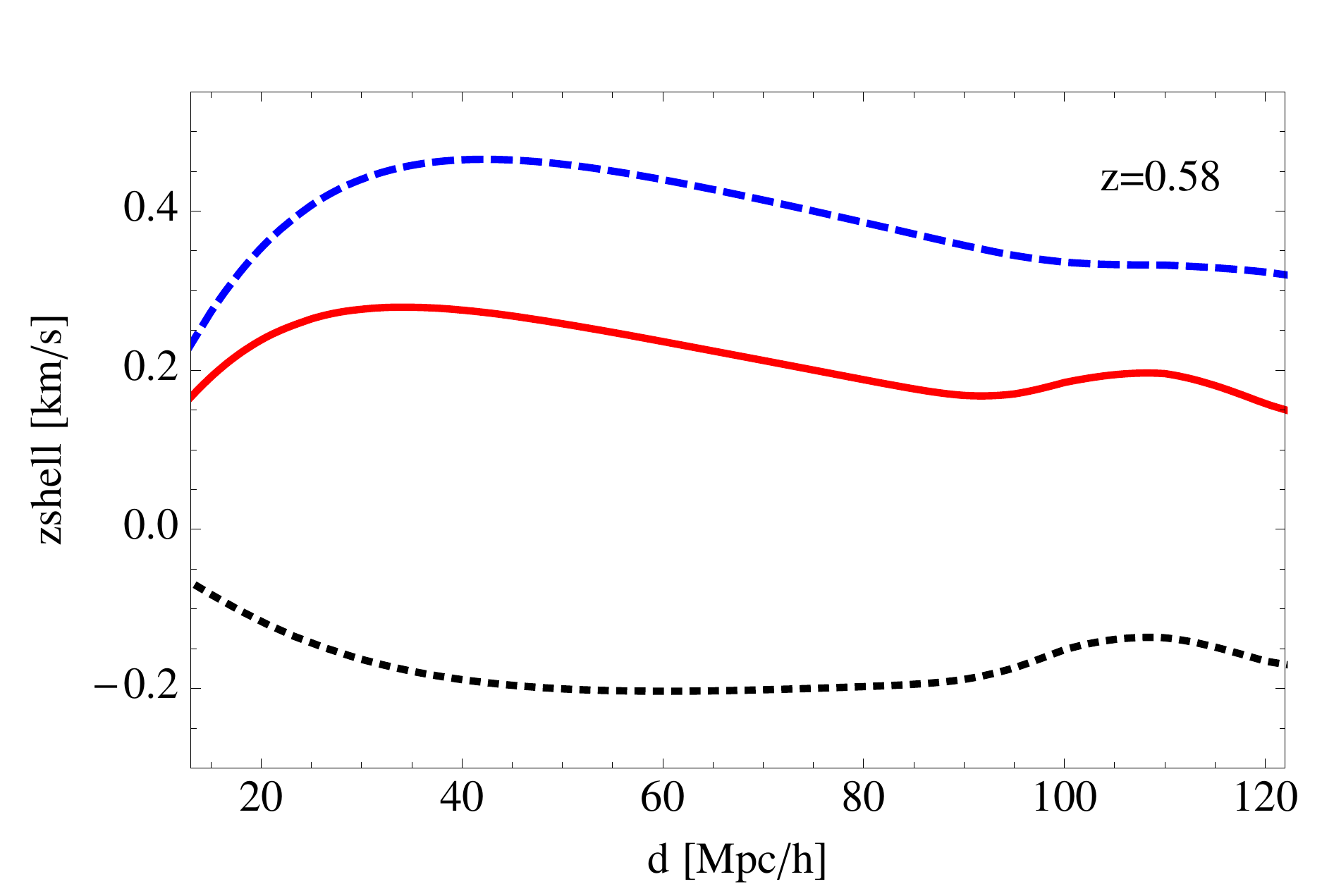}\hspace{0.8cm}\includegraphics[width=0.47\textwidth]{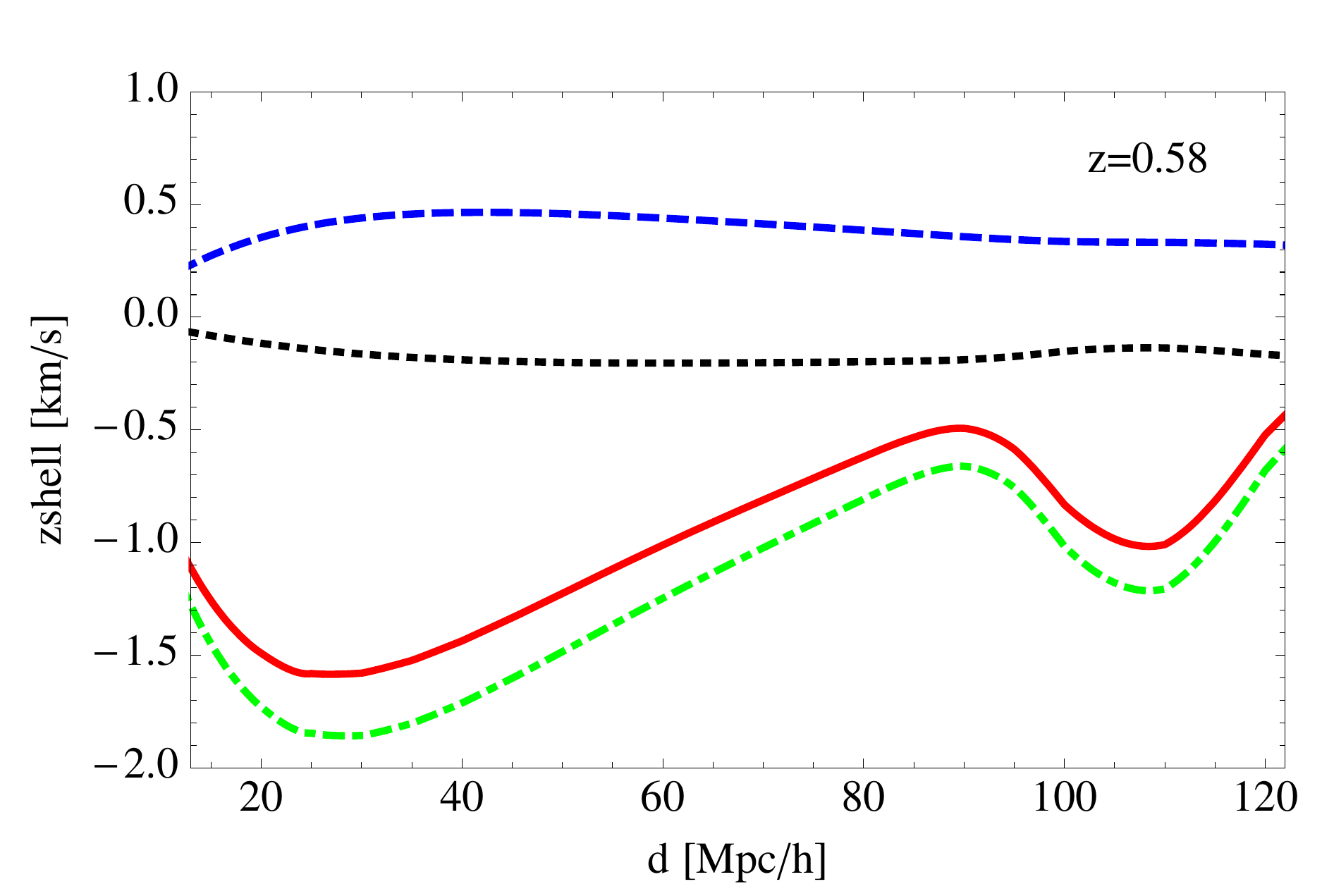}
\caption{\label{fig:zshell} Predictions for the estimator $z_{\rm shell}$ at $z=0.30$ and $z=0.58$, plotted as a function of separation. In the left panels we show $z_{\rm shell}$ calculated with the angle $\mu_{12}=\cos\sigma$ and in the right panels with $\mu_{1}=\cos\beta$. The blue dashed line is the relativistic contribution, the black dotted line is the wide-angle contribution, the green dot-dashed line is the large-angle contribution and the red solid line is the total. }
\end{figure}

\begin{figure}
\centering
\includegraphics[width=0.47\textwidth]{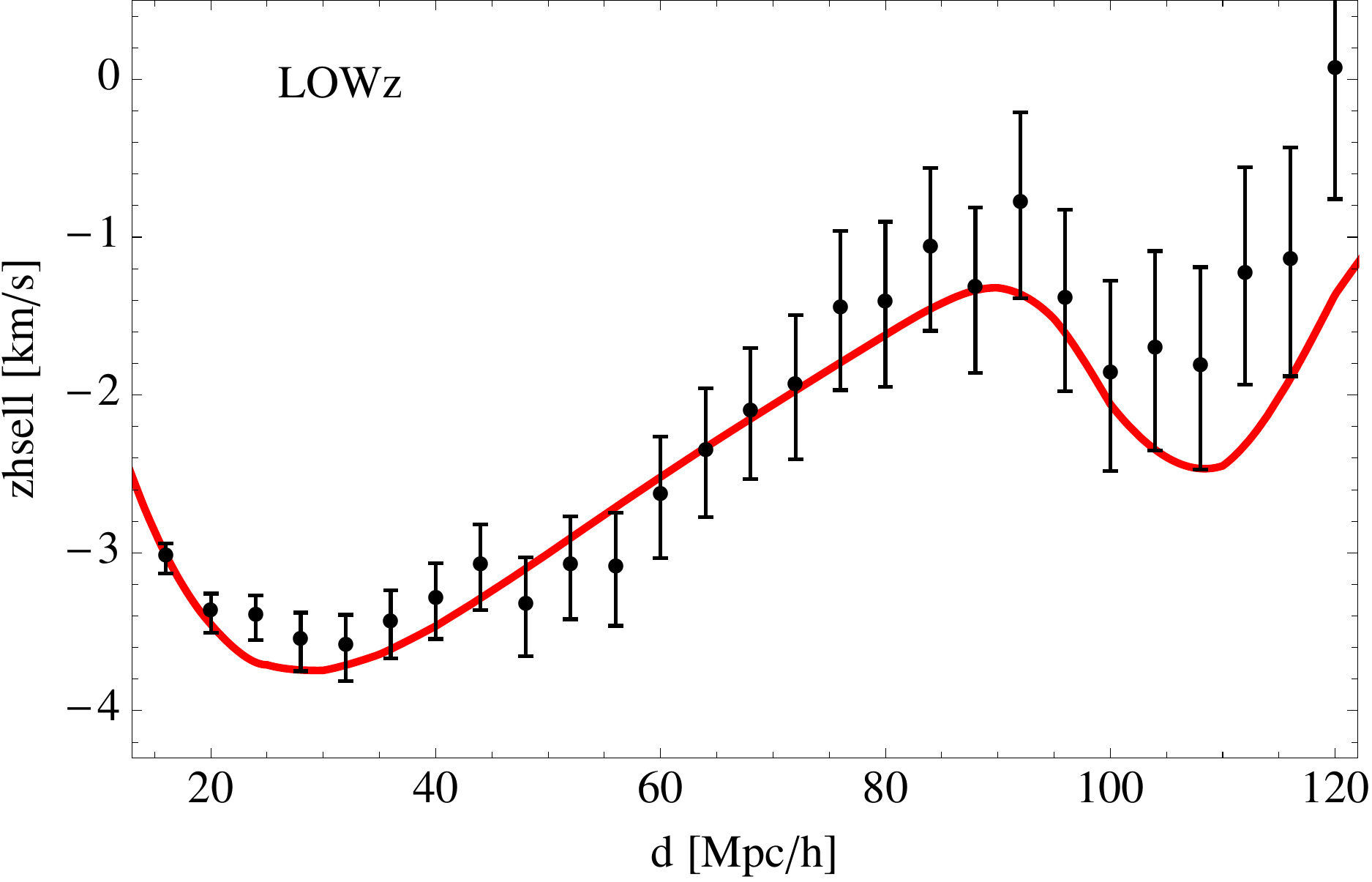}\hspace{0.8cm}\includegraphics[width=0.47\textwidth]{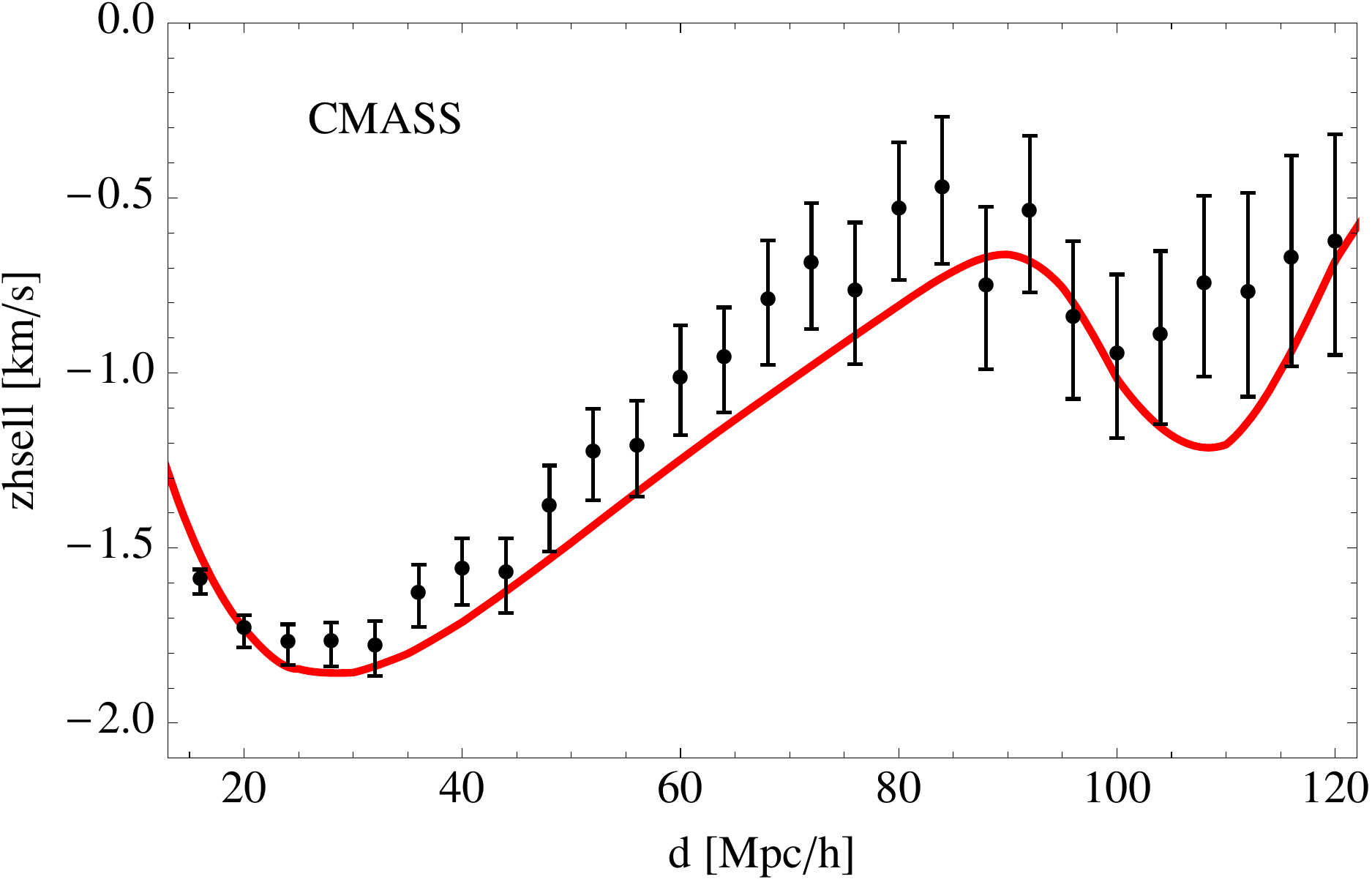}
\caption{\label{fig:zshell_BOSS} Measurements of $z_{\rm shell}$ using $\mu_1-\mu_{12}=\cos\beta-\cos\sigma$, from the cross-correlation of bright and faint galaxies in the LOWz and CMASS samples, plotted as a function of separation. The red solid line is the prediction using linear perturbation theory and the biases $b_\B$ and $b_\F$ are fitted from the total and cross monopoles, Eqs.~\eqref{monoT} and~\eqref{mono}.}
\end{figure}

\subsection{Comparison with the estimator $z_{\rm shell}$}
\label{sec:zshell}

In~\cite{Croft:2013taa}, Croft constructed an estimator $z_{\rm shell}$ to measure the redshift difference generated by gravitational redshift in large-scale structure. Writing the parallel separation between galaxies as $s_\parallel=s\cos{\gamma}$ and integrating over the orientation of the pair $\mu=\cos\gamma$, Eq.~(9) in~\cite{Croft:2013taa} can be rewritten as
\begin{align}
\label{zshell}
z_{\rm shell}=\frac{\int_{-1}^1d\mu \int_d^{d+\Delta d}ds\, s^2 H s_\parallel\zeta(s,\mu)}{\int_{-1}^1d\mu \int_d^{d+\Delta d}ds\, s^2 (1+\zeta(s,\mu))}=
\frac{\int_{-1}^1d\mu \int_d^{d+\Delta d}ds\, s^3 \mu H\zeta(s,\mu)}{\int_{-1}^1d\mu \int_d^{d+\Delta d}ds\, s^2 (1+\zeta(s,\mu))}=\frac{1}{3}\frac{ \int_d^{d+\Delta d}ds\, s^3 H\zeta_1(s)}{\int_d^{d+\Delta d}ds\, s^2 (1+\zeta_0(s))}\, ,
\end{align}
where $H=\HH/a$ is the Hubble parameter, $\zeta$ is the two-point correlation function of galaxies and $\zeta_0$ and $\zeta_1=\langle\hat\xi\rangle$ denote respectively the monopole and dipole contributions. Here we see that the angle $\beta$ is the most natural choice to measure $z_{\rm shell}$. Eq.~\eqref{zshell} can indeed be interpreted in the following way: one picks up a first galaxy in the sample, correlates this galaxy with all galaxies within a shell $d\leq s\leq d+\Delta d$, weights each pair by the estimated redshift separation $\Delta z=H s_\parallel$, and averages over all galaxies in the shell. The procedure is then repeated for all galaxies in the sample. In this procedure the parallel separation $s_{\parallel}$ is naturally defined with respect to the angle centred on the first galaxy $s_{\parallel}=s\cos\beta$.

For $\Delta d$ sufficiently small, Eq.~\eqref{zshell}can be simplified, assuming that the integrand is constant over the shell and we find
\be
z_{\rm shell}\simeq \frac{1}{3}\frac{d H(\bar z)\langle\hat\xi\rangle(d)}{1+\zeta_0(d)}\, .
\ee
The estimator $z_{\rm shell}$ can therefore easily be calculated using our expression for the dipole. In Figure~\ref{fig:zshell} we show the predictions for the different contributions to $z_{\rm shell}$ at $z~=~0.30$ and $z=0.58$. In the right panel we use the natural choice of angle $\mu_1=\cos\beta$, whereas in the left panel we use the symmetric angle $\mu_{12}=\cos\sigma$. We see that the relativistic distortions generate a redshift difference $z_{\rm shell}$ of the order of 1\,km/s at low redshift and 0.4\,km/s at high redshift. 
The wide-angle contribution is slightly smaller: of the order of -0.5\,km/s at low redshift and -0.2\,km/s at high redshift. The large-angle effect on the other hand produces a significant redshift difference, of the order of -4\,km/s at low redshift and -2\,km/s at high redshift. In Figure~\ref{fig:zshell_BOSS} we show measurements of $z_{\rm shell}$ in the LOWz and CMASS samples using $\mu_{1}-\mu_{12}$ to isolate the large-angle effect. As for the dipole, a significant detection of the large-angle effect in $z_{\rm shell}$ is possible due to the cancellation of the flat-sky part of the error.

This measurement of the large-angle effect in the estimator $z_{\rm shell}$ should be distinguished from the measurements of \cite{Wojtak:2011ia, Sadeh:2014rya} using clusters of galaxies. Ref. \cite{Wojtak:2011ia} measured the redshift difference between the brightest galaxies at the centre of clusters and the other members of the clusters in the SDSS data release DR7. They found a mean difference of $-7.7 \pm 3$\,km/s. Ref. \cite{Sadeh:2014rya} performed a similar analysis in the SDSS data release DR10, and found a mean redshift difference of $-11 + 7 -5$\,km/s. These measurements have been compared with the mean redshift difference predicted by gravitational redshift and found to be consistent (note however the discussion in~\cite{2013PhRvD..88d3013Z, Kaiser:2013ipa}, which emphasises the importance of non-linear velocity contributions).
From Figure~\ref{fig:zshell} and~\ref{fig:zshell_BOSS}, we see that with the estimator $z_{\rm shell}$ we are able to measure the large-angle effect but not the relativistic contribution (which includes the gravitational redshift effect). The relativistic contribution to $z_{\rm shell}$ is indeed of the order of $1$\,km/s in the LOWz sample, i.e. $\sim10$ times smaller than the redshift difference measured in clusters. It has also the opposite sign with respect to the redshift difference in clusters.

As explained in~\cite{Bonvin:2013ogt} (see Appendix E), this difference between the measurements from clusters and the ones from large-scale structure comes from the fact that clusters do have a physical boundary. This boundary is used in the measurement of the redshift difference, which is averaged from the minimal redshift of the cluster $z_{\rm min}$ to the maximal redshift of the cluster $z_{\rm max}$. The dominant effect in the measurement comes then from the asymmetric shift in the cluster's boundaries with respect to the centre due to gravitational redshift: $z_{\rm max}-z_\B<z_\B-z_{\rm min}$, where $z_\B$ is the redshift of the brightest galaxy in the cluster. As shown in~\cite{Bonvin:2013ogt}, this effect is in reality a three-point correlation function $\langle\Delta_\B \Delta_\F \partial_r \Psi\rangle$ and it is negative. On the other hand, in large-scale structure we do not have a natural boundary within which we can integrate the estimator $z_{\rm shell}$. Therefore the average is performed over a symmetric boundary in redshift-space: in~\eqref{zshell} the two-point function is averaged within a symmetric shell of width $\Delta d$ situated at distance $d$ from the central galaxy. The effect which is measured through $z_{\rm shell}$ (and similarly through the dipole) is therefore the effect of gravitational redshift on the two-point correlation function directly and not on the boundary of the shell. From Figure~\ref{fig:zshell} we see that this effect is much smaller than the boundary effect and of opposite sign. This explains why it is more difficult to measure gravitational redshift in large-scale structure than in clusters. 

Let us finally mention that the measurement of gravitational redshift in clusters is probably also affected at some level by wide-angle and large-angle effects. The redshift difference $z_\B-z_\F$ is indeed averaged over galaxies at fixed transverse separation from the centre. This transverse separation is naturally calculated with the angle $\beta$, centred on the bright galaxy, and one can therefore expect both a wide-angle contribution (which is probably very small at the scale of a cluster) and a large-angle contribution. Since the form of the estimator used in clusters measurement is different from $z_{\rm shell}$ and from the dipole (in particular, in clusters the weight $\cos\beta$ is not used), it is not trivial to estimate the importance of the large-angle effect in this case.

\subsection{Large-angle dipole in a single population of galaxies}
\label{sec:1pop}

The relativistic, evolution and wide-angle dipoles all vanish in the case where we have only one population of galaxies. This is clearly visible from Eqs.~\eqref{rel},~\eqref{evol} and~\eqref{wide1} which show that these effects are sensitive to the bias difference between the two populations of galaxies. The large-angle dipole on the other hand does not depend on the bias difference between the populations, see Eq.~\eqref{large}.  As such this effect should also exist in a single population of galaxies.

\begin{figure}
\centering
\includegraphics[width=0.45\textwidth]{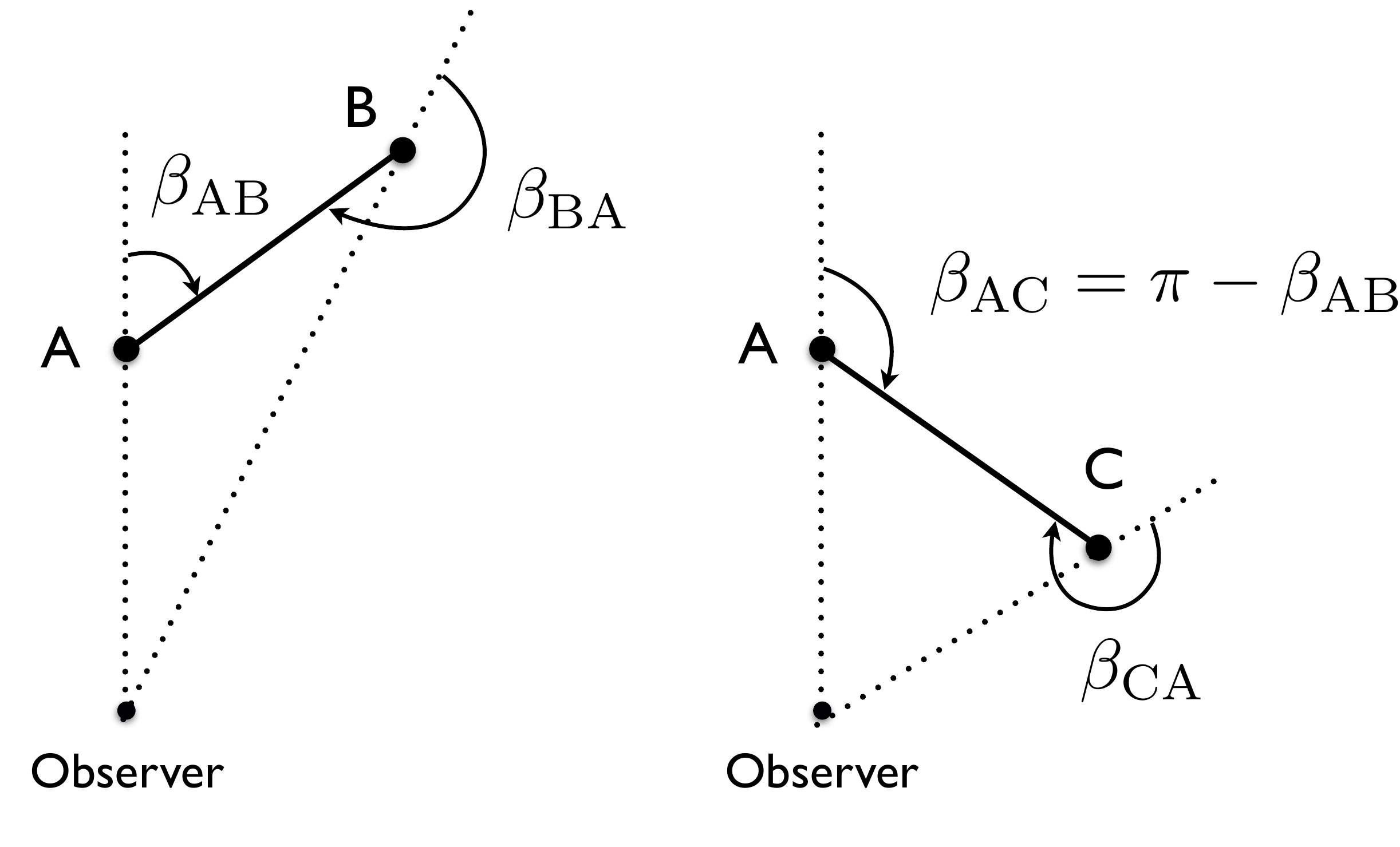}
\caption{\label{fig:pair} The contribution AB cancels with the contribution AC, since $\cos\beta_{\rm AC}=-\cos\beta_{\rm AB}$. However at large separation, the contribution CA does not cancel with the contribution BA, since  $\cos\beta_{\rm CA}\neq-\cos\beta_{\rm BA}$.}
\end{figure}

\begin{figure}
\centering
\includegraphics[width=0.47\textwidth]{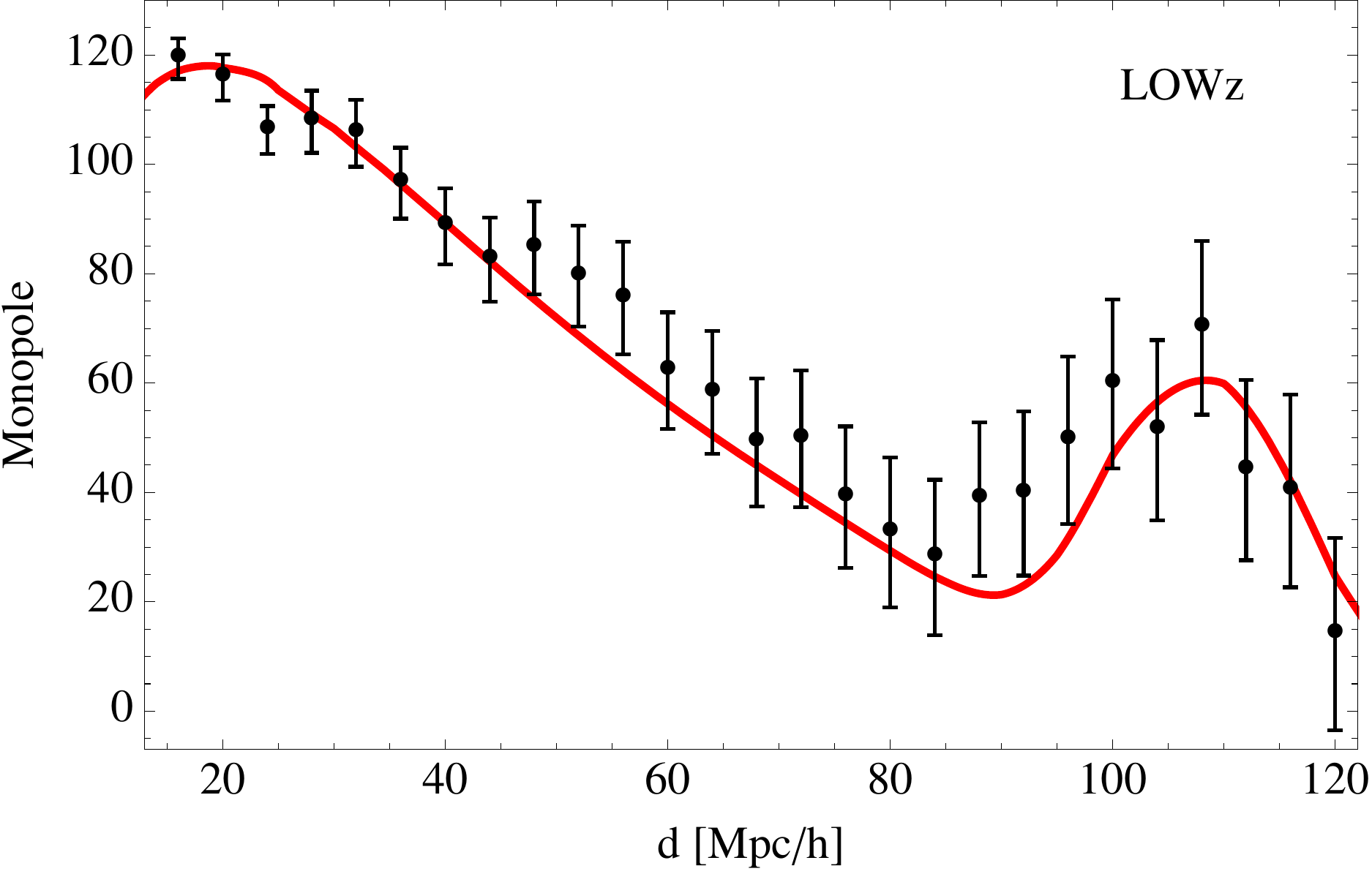}\hspace{0.8cm}\includegraphics[width=0.47\textwidth]{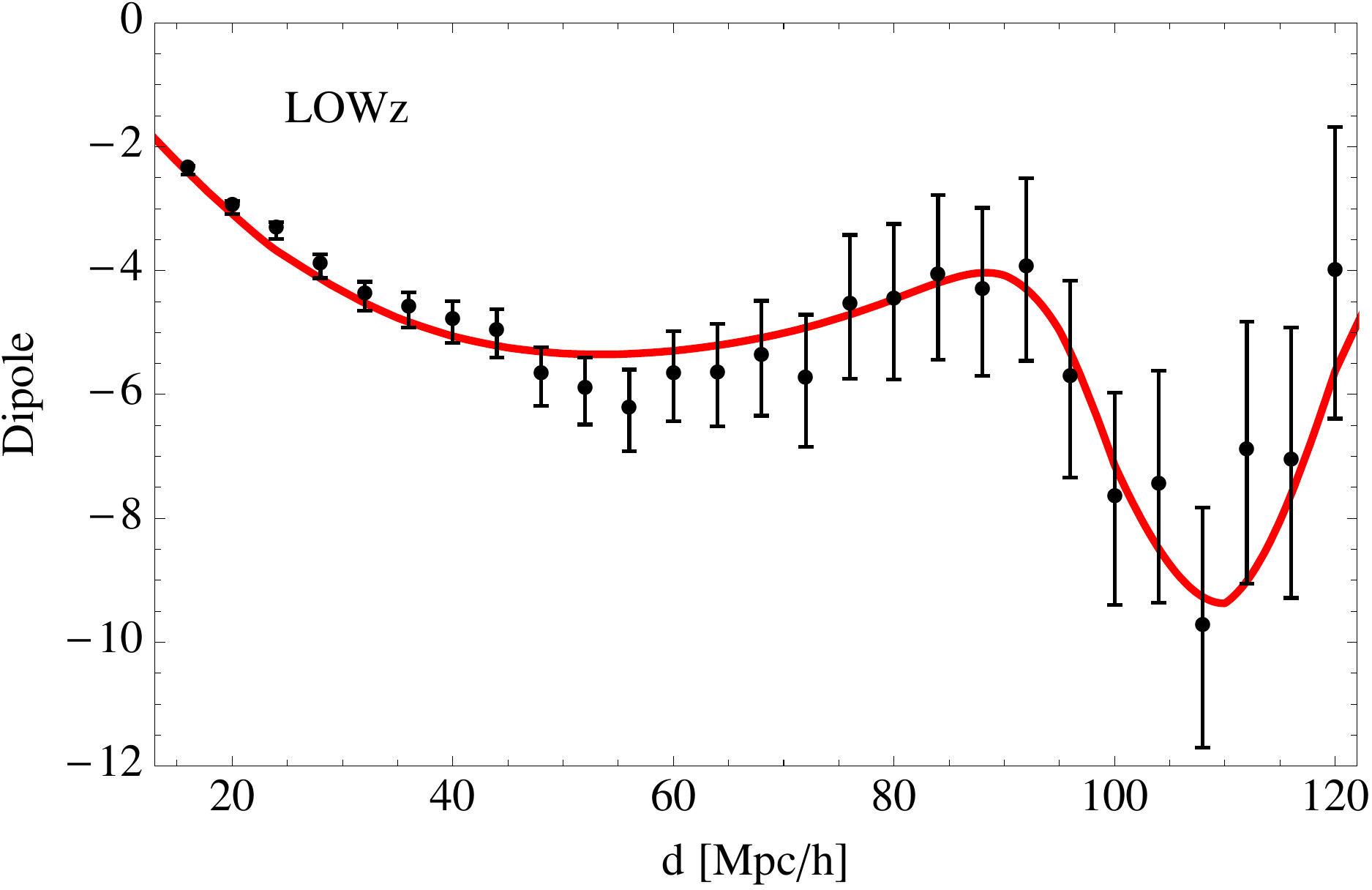}\\\includegraphics[width=0.47\textwidth]{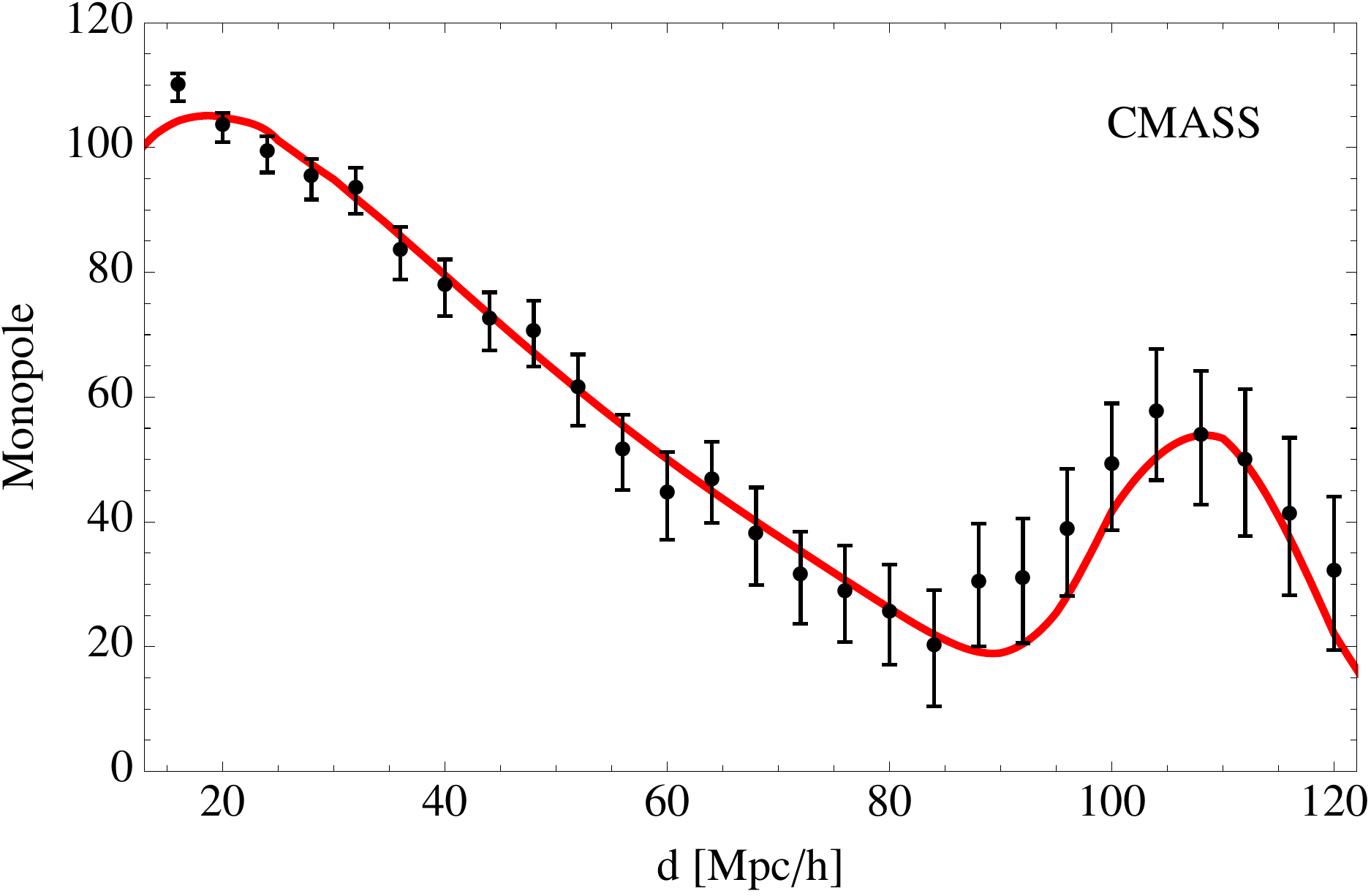}\hspace{0.8cm}\includegraphics[width=0.47\textwidth]{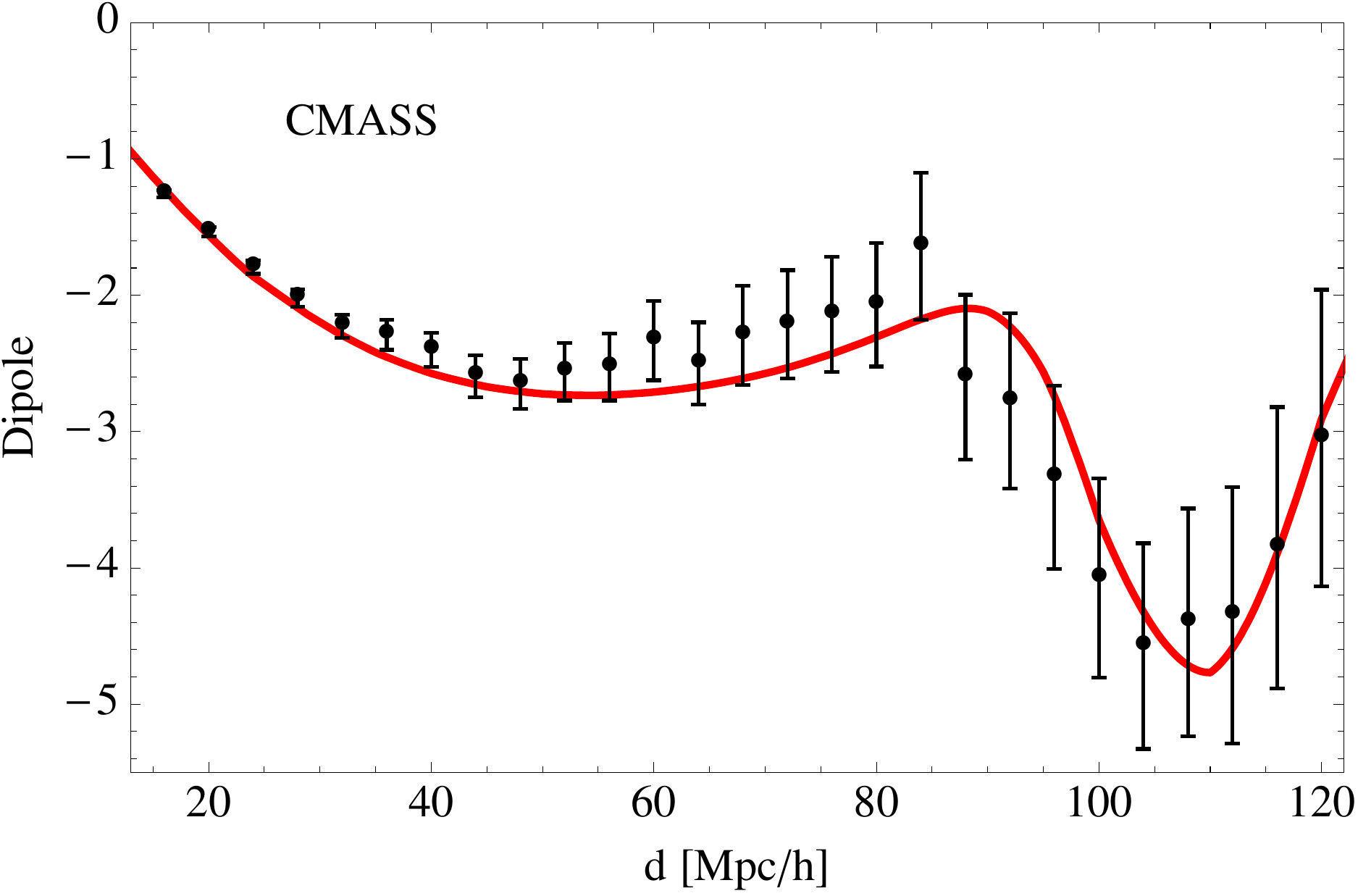}
\caption{\label{fig:1pop} Measurements of the monopole and dipole with one population of galaxies, using the difference of angles $\mu_1-\mu_{12}=\cos\beta-\cos\sigma$, multiplied by $d^2$ and plotted as a function of separation. The red solid line is the prediction using linear perturbation theory and the bias is fitted from the monopole.}
\end{figure}

\begin{figure}
\centering
\includegraphics[width=0.47\textwidth]{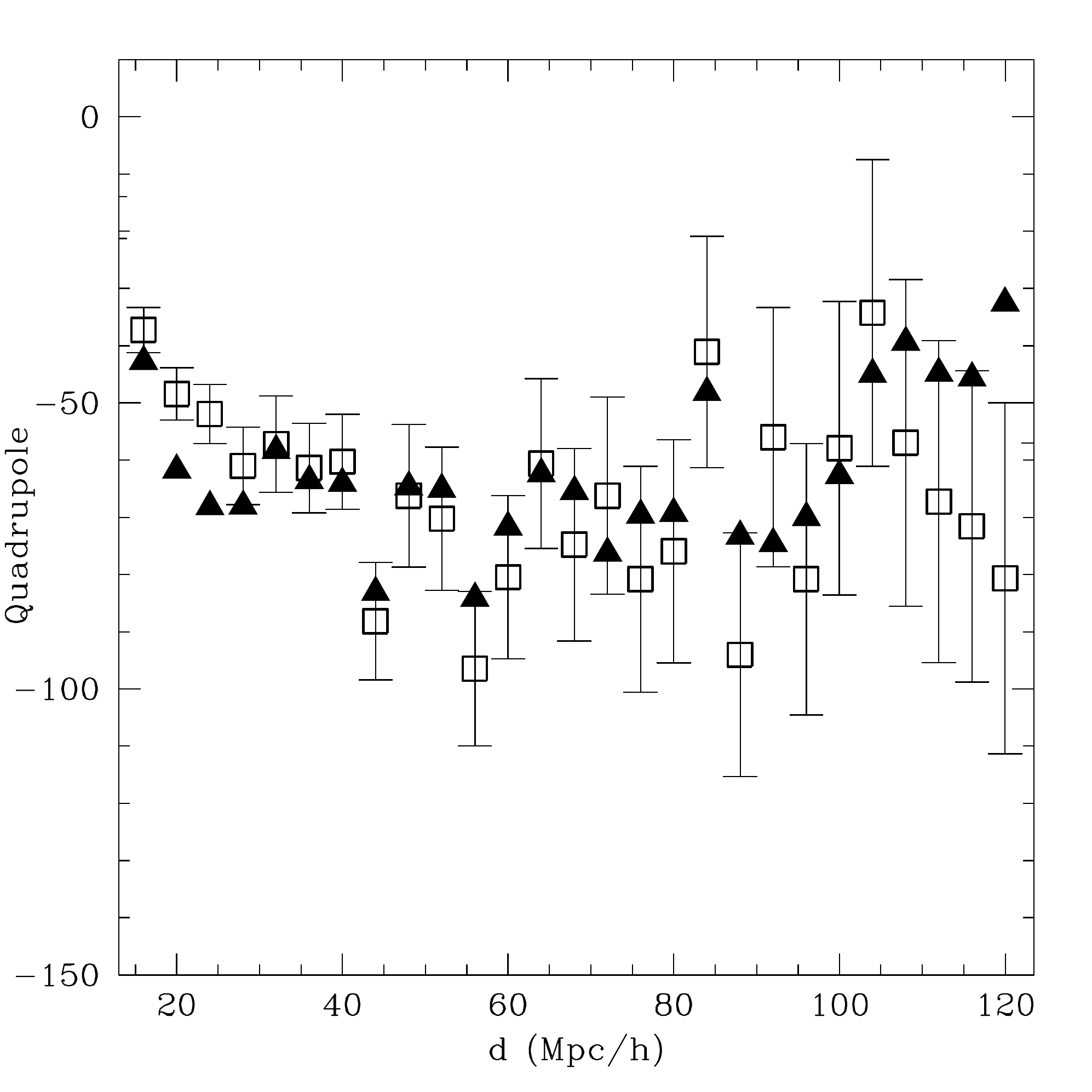}
\caption{\label{fig:dipquad} Comparison of the quadrupole measured directly from the two-point correlation function (squares with error bars) with the quadrupole extracted from the combined measurement of the monopole and the dipole using Eq.~\eqref{dip1pop} (triangles). The error bars on the triangles are similar to the one on the squares and therefore we do not plot them here for clarity.}
\end{figure}

\begin{figure}
\centering
\includegraphics[width=0.44\textwidth]{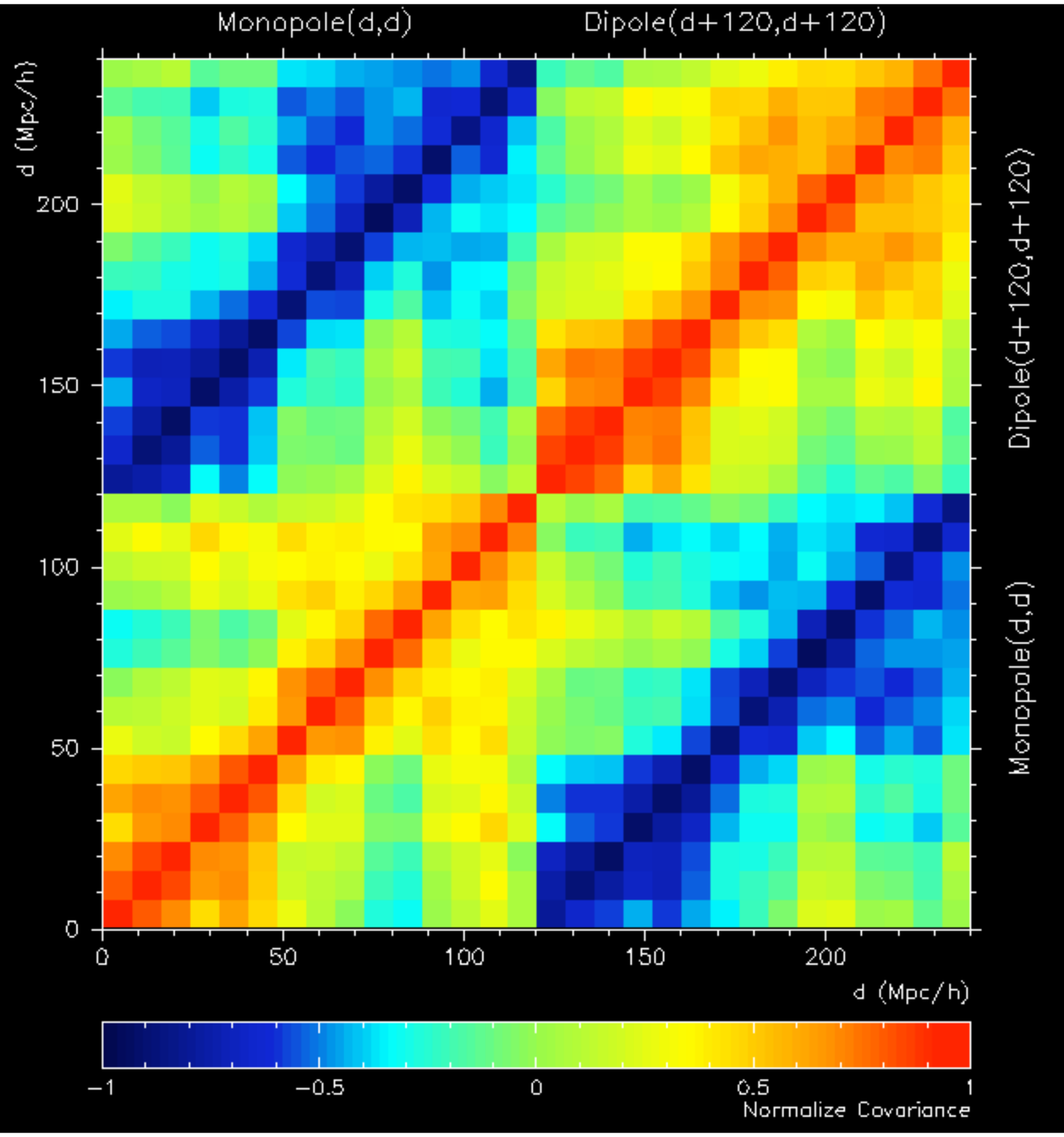}
\caption{\label{fig:cov} Covariance between the monopole and the dipole.}
\end{figure}

With one population of galaxies, we can construct the following estimator
\be
\label{xi1pop}
\hat\xi=\norm\sum_{ij} \delta n(\bx_i)\delta n(\bx_j)\cos\gamma_{ij}\delta_K(d_{ij}-d)\, .
\ee
Since there is no distinction between the two galaxies in the pair we always define the angle $\gamma_{ij}$ with respect to the first pixel $i$. Each pair will automatically be counted twice: for example a pair A-B will contribute as
\be
\xi_{\rm AB}=\norm d\bar n^2\Delta_{\rm A}\Delta_{\rm B}\left(\cos\gamma_{\rm AB}+\cos\gamma_{\rm BA}\right)\delta_K(d_{\rm AB}-d)\, ,
\ee
where in $\Delta_{\rm A}$ and $\Delta_{\rm B}$ we include only the standard density and redshift-space distortions since we know that the relativistic effects vanish for one population (at lowest order in $d/r$). 
From Figure~\ref{fig:angles} we see that if we choose $\gamma=\sigma$ then $\sigma_{\rm BA}=\sigma_{\rm AB}+\pi$ so that
\be
\cos\sigma_{\rm AB}+\cos\sigma_{\rm BA}=0\, ,
\ee
and the contribution from {\it each} pair exactly vanishes. If on the other hand we choose $\gamma=\beta$, then $\cos\beta_{\rm BA}\neq -\cos\beta_{\rm AB}$ and the contribution does not vanish. From Eq.~\eqref{betaangle} we obtain
\be
\xi_{\rm AB}=-\norm d\bar n^2 \Delta_{\rm A}\Delta_{\rm B}\frac{d_{\rm AB}}{r}\left(1-\cos^2\beta_{\rm AB}\right)\delta_K(d_{\rm AB}-d)\, .
\ee
Summing over all pairs (i.e. integrating over all orientations), we obtain then
\be
\label{dip1pop}
\langle \hat\xi \rangle=\langle \hat\xi^{\rm large} \rangle=-\frac{d}{r}\left[\zeta_0(d)-\frac{1}{5}\zeta_2(d)\right]\, .
\ee
This result may seem counter-intuitive: one could indeed argue that since we multiply by a cosine, for each pair A-B there exist another pair A-C with an opposite cosine which should cancel with A-B. This is however not correct at large separation. From Figure~\ref{fig:pair} we see that if we choose C such that $\cos\beta_{\rm AC}=-\cos\beta_{\rm AB}$ we remove only the first contribution to the pair. We have indeed
\begin{align}
\xi_{\rm AB}+&\xi_{\rm AC}=\norm d\bar n^2\Big[\Delta_{\rm A}\Delta_{\rm B}\big(\cos\beta_{\rm AB}+\cos\beta_{\rm BA}\big)\delta_K(d_{\rm AB}-d)+\Delta_{\rm A}\Delta_{\rm C}\big(\cos\beta_{\rm AC}+\cos\beta_{\rm CA}\big)\delta_K(d_{\rm AC}-d)\Big]\, . \label{pairs}
\end{align}
Since at fixed separation $d$, $\Delta_{\rm A}\Delta_{\rm B}=\Delta_{\rm A}\Delta_{\rm C}$, the first term in the first line of Eq.~\eqref{pairs} cancels with the first term in the second line. However the second terms do not cancel because
\be
\cos\beta_{\rm BA}+ \cos\beta_{\rm CA}=-2\frac{d}{r}\left(1-\cos^2\beta_{\rm AB}\right)\neq 0\, .
\ee
Therefore in the full-sky there is no pair which exactly cancels the contribution from the pair A-B. Note that the existence of a large-angle dipole in one population of galaxies has already been demonstrated in~\cite{2016JCAP...01..048R}.

We measure the large-angle dipole with one population in the LOWz and CMASS samples. 
As for two populations, we use the combination of angle $\cos\beta-\cos\sigma$ to remove the dominant part of the cosmic variance (see Section~\ref{sec:case 3}). 
The results are plotted in Figure~\ref{fig:1pop}. 
We see that the dipole agrees very well with the predictions both in the LOWz and in the CMASS samples. 

The large-angle dipole could provide an alternative way of measuring redshift-space distortions with different systematics. In Figure~\ref{fig:dipquad} we compare the quadrupole extracted from the dipole using Eq.~\eqref{dip1pop} with the quadrupole measured from the two-point function in the standard way. The two quadrupoles are not exactly equivalent, which suggests that they are affected by systematics in a different way. Figure~\ref{fig:cov} shows the covariance between the dipole and the monopole. These two multipoles are not perfectly anti-correlated, showing that they do not contain exactly the same information. The dipole is therefore potentially useful to measure redshift-space distortions.

Finally, let us mention the following point: from Eq.~\eqref{dip1pop} it seems that by combining the measurements of the monopole, the quadrupole and the large-angle dipole, we can directly measure the comoving distance $r$
\be
\label{r1pop}
r= -\frac{d}{\langle\hat\xi^{\rm large} \rangle}\left[\zeta_0(d)-\frac{1}{5}\zeta_2(d)\right]\, .
\ee
However the distance measured in this way is not the true distance but rather the distance in the fiducial cosmology used to measure the multipoles. The relation between the dipole, the monopole and the quadrupole is indeed due to the fact that in the full sky $\cos\beta_{\rm AB}$ and  $\cos\beta_{\rm BA}$ are related through Eq.~\eqref{betaangle}. This relation is a geometrical relation which is valid in {\it any} fiducial cosmology. Therefore contrary to the position of the BAO peak, this relation does not allow us to test the validity of the fiducial cosmology. Note however that we can use the dipole to measure the position of the BAO peak, in a similar way as is done with the monopole, see figures~\ref{fig:dipole3} and~\ref{fig:1pop}.

\section{Conclusions}
\label{sec:conclusion}

We have presented a first attempt to measure the dipole in the cross-correlation function of bright and faint galaxies in the LOWz and CMASS samples of the BOSS survey. We have identified four types of contributions to the dipole: relativistic distortions, evolution effect, wide-angle effect and large-angle effect. We have shown that the first three effects generate intrinsic anti-symmetric contributions that exist for any choice of angle used to measure the dipole. The large-angle effect on the other hand appears only if the angle chosen to extract the dipole breaks the symmetry of the problem.

We have found that the dipole from relativistic distortions, evolution effect and wide-angle effect is too small to be detected in BOSS since its signal-to-noise is of the order of 0.2. On the other hand we have measured the large-angle dipole using the combination of angles $\mu_1-\mu_{12}=\cos\beta-\cos\sigma$. This large-angle dipole does not contain any new physical information, since it is just a geometrical combination of the monopole and the quadrupole. However, this detection is interesting for various reasons: first it shows that if one wants to measure relativistic effects in future surveys it is crucial \textit{not} to use the angle $\mu_{12}=\cos\beta$. This angle seems the most natural one to measure an effect like gravitational redshift, but it has the strong disadvantage to artificially introduce anti-symmetries in the correlation function. Second, since the measurement of the large-angle dipole is in very good agreement with our theoretical prediction, it validates our method for extracting the dipole from the two-point correlation function. Finally, 
it is conceivable that the large-angle dipole is sensitive to different systematics than the monopole and the quadrupole. For example since the kernel to isolate the dipole is anti-symmetric, all types of systematics which are symmetric will automatically cancel out from the dipole. The large-angle dipole could therefore be used as an alternative to the quadrupole to measure redshift-space distortions.

Our analysis shows that the relativistic distortions, which include the effect of gravitational redshift, are not detectable in the BOSS survey. Comparing our predictions with the measurement of gravitational redshift in clusters, we find that the signal in BOSS is 10 times smaller. This difference is likely due to the fact that in clusters one has additional information-- the boundaries of the clusters-- which can be used to measure the gravitational redshift effect. This however does not mean that the relativistic effects cannot be detected at all in large-scale structure. In~\cite{Bonvin:2015kuc}, we showed how to construct an optimal estimator to measure the relativistic dipole with an arbitrary number of galaxy populations. We predicted that in the main sample of the SDSS data release DR5, using 6 populations, the cumulative signal-to-noise of the relativistic dipole reaches 2.4. With the up-comming  DESI survey, which will contain significantly more galaxies than SDSS we predict a cumulative signal-to-noise of 7.4 for the relativistic dipole, showing that our method could readily be used to detect gravitational redshift in large-scale structure in the near future.

\[\]
{\bf Acknowledgments:}  It is a pleasure to thank Rupert Croft, Alex Hall, Ofer Lahav and Francesco Montanari for interesting and useful discussions.
EG acknowledges support from projects AYA2012-39559, AYA2015-71825-P and  Consolider-Ingenio CSD2007- 00060 
from he Spanish Ministerio de Ciencia e Innovacion (MICINN) and CosmoComp (PITNGA-2009-238356) and 
LACEGAL (PIRSES-GA-2010-269264) from the European Commission. CB acknowledges support by the Swiss National Science Foundation. LH is support in part by the DOE and NASA.

\appendix

\section{Other choice of kernel}
\label{app:kernel}

In~\cite{Bonvin:2015kuc} we use a slightly different choice of kernel than in Eq.~\eqref{kernel2}. Instead of defining the angle  $\cos\gamma_{ijL_i L_j}$ with respect to the bright galaxy in the pair, as done in~\eqref{gamma}, we define it with respect to the first pixel in the pair, i.e. $i$. We then multiply the pair of pixels by $+1$ if the pixel $i$ is bright and the pixel $j$ is faint, and by $-1$ if the pixel $i$ is faint and the pixel $j$ is bright. Such a kernel can be written as
\be
\label{kernel1}
\w{i}{j}=\norm\Big[\theta(L_i-L_j) -\theta(L_j-L_i) \Big] \cos\gamma_{ij}\delta_K(d_{ij}-d)\, ,
\ee
where  $\theta$ is the Heaviside function and $\norm$ is defined in Eq.~\eqref{norm}.

The relativistic dipole and the evolution dipole are exactly the same with this choice of kernel. They are given by Eqs.~\eqref{rel} and~\eqref{evol}.
The wide-angle dipole is given by 
\begin{align}
\label{xiw1sigma}
\langle \hat\xi^{\rm wide2\, \sigma} \rangle=\norm\dn_{\B}\dn_{\F}\sum_{ij}\cos\sigma_{ij}\delta_K(d_{ij}-d)\Big[\langle \Delta_\B(\bx_i)\Delta_\F(\bx_j)\rangle-\langle \Delta_\B(\bx_j)\Delta_\F(\bx_i)\rangle\Big]^{\rm wide}\, ,
\end{align}
if we use the angle $\sigma$ and by
\begin{align}
\label{xiw1beta}
\langle \hat\xi^{\rm wide2\, \beta} \rangle=&\norm\dn_{\B}\dn_{\F}\sum_{ij}\cos\beta_{ij}\delta_K(d_{ij}-d)\Big[\langle \Delta_\B(\bx_i)\Delta_\F(\bx_j)\rangle-\langle \Delta_\B(\bx_j)\Delta_\F(\bx_i)\rangle\Big]^{\rm wide}\, ,
\end{align}
if we use the angle $\beta$ (note that this is the case used in~\cite{Bonvin:2013ogt}). Since the bracket is already of order $d_{ij}/r$, these two expressions are equivalent and similar to Eq.~\eqref{xisigma}. We have therefore
\be
\label{wide3ways}
\langle \hat\xi^{\rm wide2\, \sigma} \rangle=\langle \hat\xi^{\rm wide2\, \beta} \rangle=\langle \hat\xi^{\rm wide\, \sigma} \rangle=\frac{2f}{5}(b_\B-b_\F)\frac{d}{r}C_2(d)\, .
\ee

To summarise, we have 4 possible choices of kernel. We have two different possibilities of choosing the angle and two different possibilities of choosing the dependence in luminosity. Three of these possibilities give exactly the same wide-angle contribution~\eqref{wide3ways}. The other possibility, which consists in choosing kernel~\eqref{kernel2} with the angle $\beta$, gives an additional contribution, the large-angle dipole, calculated in Eq.~\eqref{wide2}.

\section{Explicit calculation of the wide-angle effect in terms of the angle $\sigma$}
\label{app:wideangle}

In~\cite{Bonvin:2013ogt} we derived a general expression, valid in the full sky, for the standard part of the two-point correlation function
\begin{align}
\langle \Delta_\B(\bx_i)\Delta_\F(\bx_j)\rangle=\frac{2A}{9\pi^2\Om_m^2}\Big\{S_1+S_2\cos(2\beta_{ij}) +S_3\cos(2\alpha_{ij})
+S_4\cos(2\alpha_{ij})\cos(2\beta_{ij})+S_5\sin(2\alpha_{ij})\sin(2\beta_{ij}) \Big\}\, ,
\end{align}
where $A$ is the primordial amplitude of perturbations, $\Omega_m$ is the matter density parameter, $S_\alpha$ are coefficients defined in Appendix C of~\cite{Bonvin:2013ogt}~\footnote{The coefficients $S_2$ and $S_3$ in~\cite{Bonvin:2013ogt} have a typo: there should be a minus sign in front of $1/28$.} and $\alpha_{ij}=\beta_{ji}-\pi$. We want to rewrite $\beta_{ij}$ and $\alpha_{ij}$ as a function of $\sigma_{ij}$. From Figure~\ref{fig:angles} we have
\begin{align}
&\cos(2\beta_{ij})= \cos(2\sigma_{ij})-\frac{d_{ij}}{r}\sin(2\sigma_{ij})\sin(\sigma_{ij})+\mathcal O \left(\frac{d_{ij}}{r} \right)^2\, ,\nonumber\\
&\cos(2\alpha_{ij})= \cos(2\sigma_{ij})+\frac{d_{ij}}{r}\sin(2\sigma_{ij})\sin(\sigma_{ij})+\mathcal O \left(\frac{d_{ij}}{r} \right)^2\, ,\nonumber\\
&\sin(2\alpha_{ij}) \sin(2\beta_{ij})=\sin^2(2\sigma_{ij})+\mathcal O \left(\frac{d_{ij}}{r} \right)^2\, .
\end{align}
With this we obtain up to order $d_{ij}/r$
\begin{align}
\label{standsig}
\langle \Delta_\B(\bx_i)\Delta_\F(\bx_j)\rangle=\frac{2A}{9\pi^2\Om_m^2}\Bigg\{&S_1+S_2\Big[\cos(2\sigma_{ij})-\frac{d_{ij}}{r}\sin(2\sigma_{ij})\sin(\sigma_{ij})\Big]
+S_3\Big[\cos(2\sigma_{ij})+\frac{d_{ij}}{r}\sin(2\sigma_{ij})\sin(\sigma_{ij})\Big]\nonumber\\
&+S_4\cos^2(2\sigma_{ij})+S_5\sin^2(2\sigma_{ij}) \Bigg\}\, . 
\end{align}
The dipole part of Eq.~\eqref{standsig} is given by
\be
\langle \Delta_\B(\bx_i)\Delta_\F(\bx_j)\rangle^{\rm dip}=\frac{2A}{9\pi^2\Om_m^2}\frac{4}{5}(S_3-S_2)\frac{d_{ij}}{r}\cos\sigma_{ij}\, .
\ee
As already noted in~\cite{2016JCAP...01..048R}, in the case of one population of galaxies $S_2=S_3$ and there is no wide-angle contribution to the dipole with this choice of angle. On the other hand, with two populations of galaxies we find
\begin{align}
\langle \Delta_\B(\bx_i)\Delta_\F(\bx_j)\rangle^{\rm dip}&=-\frac{2f}{5}(b_\B-b_\F)\frac{d_{ij}}{r}C_2(d_{ij})\cos\sigma_{ij}=-\langle \Delta_\F(\bx_i)\Delta_\B(\bx_j)\rangle^{\rm dip}\, .
\end{align}
Inserting this into Eq.~\eqref{xisigma} we obtain Eq.~\eqref{wide1}.


\bibliography{bibweights.bib}

\end{document}